\documentclass[sn-basic,Numbered,iicol]{sn-jnl}

\usepackage{graphicx}%
\usepackage{multirow}%
\usepackage{amsmath,amssymb,amsfonts}%
\usepackage{amsthm}%
\usepackage{mathrsfs}%
\usepackage[title]{appendix}%
\usepackage[table]{xcolor}%
\usepackage{textcomp}%
\usepackage{manyfoot}%
\usepackage{booktabs}%
\usepackage[ruled,vlined,linesnumbered]{algorithm2e}
\usepackage{listings}%
\usepackage[most]{tcolorbox}
\usepackage[T1]{fontenc}
\usepackage{cleveref}

\SetCommentSty{mycommfont}

\newcommand{\approach}{{\fontfamily{cmss}\selectfont \mbox{EnvDT}}}

\hypersetup{pdfborder={0 0 0}}

\begin{document}

\title[Digital Twins Environment Simulation]{Uncertainty-Aware Environment Simulation of Medical Devices Digital Twins} 

\author*[1]{\fnm{Hassan} \sur{Sartaj}}\email{hassan@simula.no}

\author[1]{\fnm{Shaukat} \sur{Ali}}\email{shaukat@simula.no}

\author[2]{\fnm{Julie} \sur{Marie Gjøby}}\email{julie-marie.gjoby@hel.oslo.kommune.no}

\affil*[1]{\orgname{Simula Research Laboratory}, \orgaddress{\city{Oslo}, \country{Norway}}}

\affil[2]{\orgdiv{Welfare Technologies Section}, \orgname{Oslo Kommune Helseetaten}, \orgaddress{\city{Oslo}, \country{Norway}}}


\abstract{
Smart medical devices are an integral component of the healthcare Internet of Things (IoT), providing patients with various healthcare services through an IoT-based application. 
Ensuring the dependability of such applications through system and integration-level testing mandates the physical integration of numerous medical devices, which is costly and impractical. 
In this context, digital twins of medical devices play an essential role in facilitating testing automation.
Testing with digital twins without accounting for uncertain environmental factors of medical devices leaves many functionalities of IoT-based healthcare applications untested. 
In addition, digital twins operating without environmental factors remain out of sync and uncalibrated with their corresponding devices functioning in the real environment.
To deal with these challenges, in this paper, we propose a model-based approach (\approach{}) for modeling and simulating the environment of medical devices' digital twins under uncertainties. 
We empirically evaluate the \approach{} using three medicine dispensers, Karie, Medido, and Pilly connected to a real-world IoT-based healthcare application.  
Our evaluation targets analyzing the coverage of environment models and the diversity of uncertain scenarios generated for digital twins. 
Results show that \approach{} achieves approximately 61\% coverage of environment models and generates diverse uncertain scenarios (with a near-maximum diversity value of 0.62) during multiple environmental simulations. 
}




\maketitle

\section{Introduction}\label{sec:intro}
An important part of healthcare Internet of Things (IoT) is diverse kinds of smart IoT medical devices. 
These devices have smart features, including remote patient monitoring, delivering healthcare services to patients, and informing concerned personnel about patients' health. 
For example, an automated medicine dispenser ensures the timely delivery of medications to patients. It continuously communicates with an IoT-based healthcare application to provide real-time updates on patients' medication status.
An IoT-based healthcare application connects healthcare service providers (such as hospitals and pharmacies), medical devices, patients, and caregivers to deliver healthcare services efficiently.

Testing an IoT-based healthcare application at the system \textcolor{black}{and integration levels}, where multiple medical devices are physically connected, is financially expensive, time-intensive, and impractical~\cite{sartaj2023testing}.
Oslo City's healthcare department encountered these challenges while developing an automated testing solution for IoT-based healthcare applications. 
\textcolor{black}{
From the literature~\cite{somers2023digital}, we observed that digital twins (DTs) have played a significant role in various testing activities (e.g., test oracles and test execution) across different safety-critical domains (e.g., aviation and healthcare). 
Therefore, to address the challenges associated with connecting physical medical devices, we introduced the concept of utilizing DTs of medical devices to streamline test execution automation~\cite{sartaj2023hita}. 
}
In this context, we proposed a model-based approach for creating and operating DTs of medicine dispensers~\cite{sartaj2024modelbased}. 
\textcolor{black}{
Our experiments with DTs have demonstrated their high fidelity to their physical counterparts, indicating their suitability as substitutes for medicine dispensers in testing scenarios. 
}

Our insights from experimenting with DTs showed that medical devices operate in a non-deterministic environment containing several uncertain factors. 
Operating device DTs in isolation from the environment results in omitting crucial device features necessary for testing IoT-based healthcare applications. 
This includes user interactions with devices, reporting device features (critical for patients), and device malfunctioning. 
For instance, a medicine dispenser rings an alarm while dispensing medicine and waits for a patient to take the medicine after pressing the confirmation button. 
If the patient fails to do so, the dispenser notifies concerned personnel about missed doses via an IoT-based healthcare application. 
Without simulating such scenarios for a device DT, device reporting features cannot be tested. 
Additionally, DTs' environment simulation is necessary for calibrating and synchronizing them with their physical counterparts. 

\textcolor{black}{
Upon analyzing the related literature, we identify several approaches across diverse areas, such as IoT~\cite{elayan2021digital,nguyen2022digital,sleuters2019digital,jiang2021digital,kirchhof2021understanding,sciullo2024relativistic,sartaj2024modelbased}, cyber-physical systems (CPS)~\cite{kirchhof2020model,dobaj2022towards,damjanovic2019open,zhou2024toward,barat2022digital,pirbhulal2024cognitive}, and smart manufacturing~\cite{somers2023digital,tao2019digital,shoukat2024smart,damjanovic2019open,llopis2023modeling}. 
These approaches focus on generating DTs, utilizing DTs in various fields, modeling and simulating the environment, and modeling uncertainty. 
Our work does not focus on generating or utilizing DTs, but rather on simulating the environment for DTs of IoT medical devices. 
Among the environment modeling and simulation approaches, these approaches target domains that differ from medical devices, such as smart offices~\cite{llopis2023modeling} and embedded systems~\cite{iqbal2015environment}. 
Adapting these approaches to medical devices would require extensive customization, first by tailoring the modeling methodology. 
Furthermore, a specialized approach would be required to specify and handle uncertainties in the tailored models. 
Due to differences in the target domain, the environmental factors and the associated uncertainties also vary significantly. 
For instance, user interactions with a medicine dispenser, such as taking medications, differ significantly from interactions with a smart office system, such as toggling room lights on or off. 
Similarly, the environmental uncertainties associated with a medicine dispenser, such as the possibility of medicine getting stuck, are unique and would not apply to domains like smart offices~\cite{llopis2023modeling} or embedded systems~\cite{iqbal2015environment}. 
Given the current state of research, simulating uncertain environments for DT of IoT medical devices remains an unexplored research area. 
}

In this window of opportunity, this paper introduces an approach (\approach{}) for modeling and simulating the environment of medical devices DTs in the presence of uncertainties.
\approach{} includes a profile for modeling the environment of medical devices and a model library for modeling environmental events.
In addition, \approach{} comprises a comprehensive modeling methodology designed to assist modelers in developing environmental models. 
We evaluate \approach{} by experimenting with three medicine dispensers, Karie~\cite{karie}, Medido~\cite{medido}, and Pilly~\cite{pilly} connected to a real-world IoT-based healthcare application, which the Oslo City healthcare department provides as a part of the experimental equipment. 
Our evaluation focuses on analyzing the coverage of environment models and the diversity of uncertain scenarios generated for the DTs of Karie, Medido, and Pilly. 
The results show that \approach{} achieved approximately 61\% coverage of environment models across various simulations for the Karie, Medido, and Pilly devices. 
Furthermore, the diversity results indicate that the overall diversity value obtained was 0.62, which is close to 1---signifying high diversity. 
These results suggest that \approach{} maintained a reasonable level of diversity within uncertain scenarios throughout each environmental simulation for every device.

The rest of the paper is structured in the following manner. \Cref{sec:bg} presents background, industrial context, and motivating example. 
\Cref{sec:approach,sec:modutils,sec:envmodels} provide an overview of the approach (\approach{}), environment modeling utilities, and modeling methodology and environment simulation, respectively. 
\Cref{sec:evaluation} provides the evaluation of the \approach{}.
\Cref{sec:insights} presents insights and lessons learned. 
\Cref{sec:relatedworks} relates our work with existing works.
\Cref{sec:conclusion} concludes this paper.

\section{Background and Motivation}\label{sec:bg}
In this section, first, we present an overview of an IoT-based healthcare application. 
Next, we discuss the context of our work and challenges. 
Lastly, we present a motivating scenario to highlight the significance of our research.

\subsection{IoT-based Healthcare Applications}
\begin{figure}[!t]
\centerline{\includegraphics[width=7.8cm, height=6.1cm, keepaspectratio]{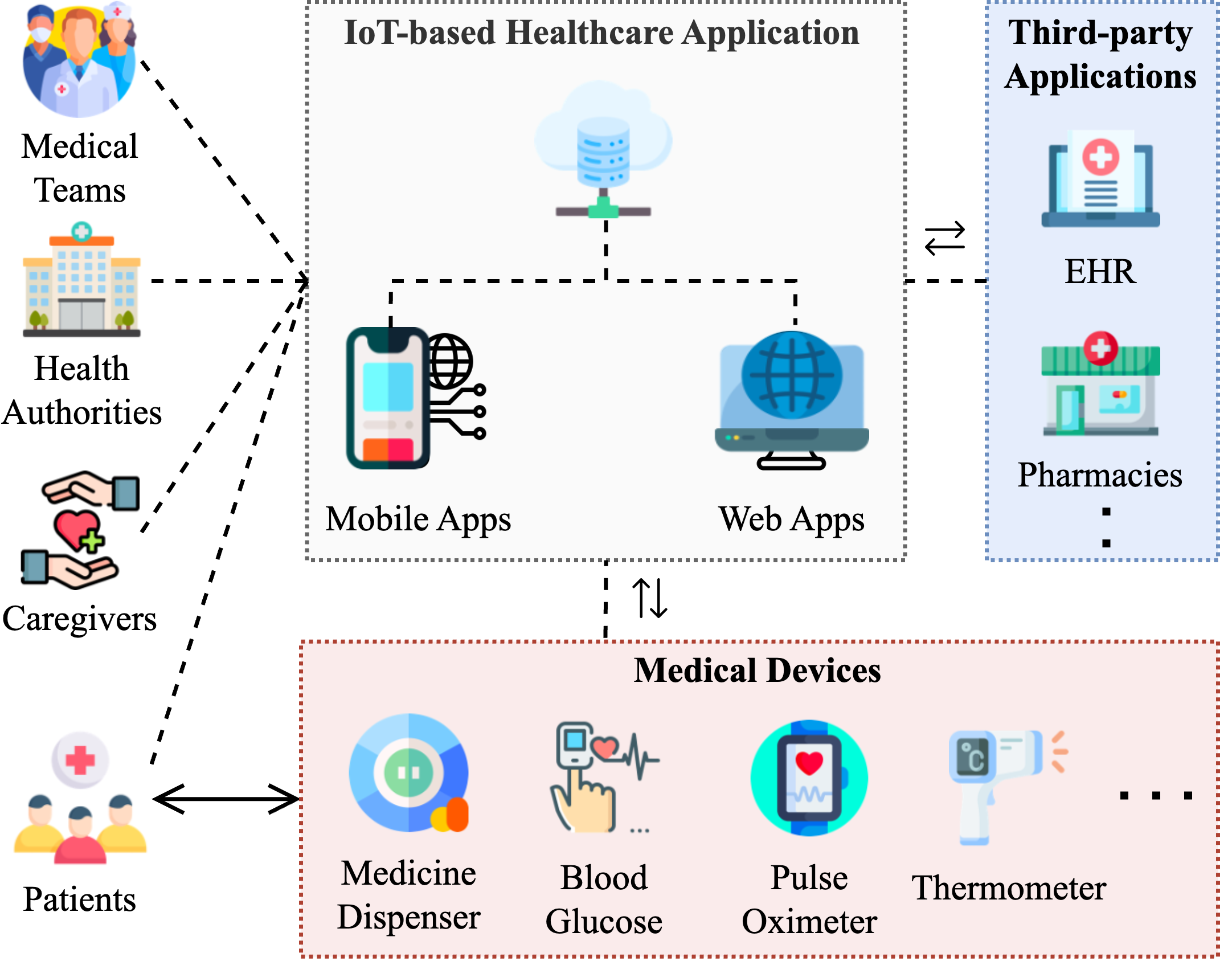}}
\caption{A common architecture of IoT-based healthcare applications.}
\label{fig:iot}
\end{figure}

\Cref{fig:iot} depicts a common architecture followed by IoT-based healthcare applications. 
Such applications utilize cloud platforms to connect various stakeholders, service providers, and medical devices. 
It has different interfaces (i.e., mobile and web applications) with specific roles for various stakeholders, mainly medical teams, health authorities, caregivers, and patients.
Under the IoT cloud platform, it connects different health service providers such as Electronic Health Records (EHR), pharmacies, hospitals, and so forth. 
Another essential part of IoT-based applications is smart medical devices of various kinds, such as medicine dispensers, blood glucose meters, pulse oximeters, and thermometers. 
These medical devices are assigned to patients, considering their health requirements. 
For example, medicine dispensers are assigned to patients for timely medicines. 
These devices not only deliver healthcare services to patients but also inform stakeholders about the patient's medical conditions, especially in emergencies.

\subsection{Context and Challenges}\label{sec:context}
Oslo City's healthcare department works with various industry partners to develop an IoT-based platform consisting of various third-party healthcare applications, smart medical devices, doctors, caregivers, and hospitals. 
The IoT-based healthcare applications developed through this process aim to provide residents with high-quality medical treatment and efficiently deliver remote healthcare services. 
In this context, medical devices play a key role. 
For instance, automated medicine dispensers deliver timely medicine to patients and constantly update an IoT-based healthcare application on the medicines consumed. 
Similarly, a pulse oximeter reports patients' pulse information, supporting remote health monitoring. 
The information sent by medical devices enables medical professionals and caregivers to analyze patients' health conditions and timely deliver medical care. 

Due to the critical nature of IoT-based healthcare applications, ensuring their dependability is a primary focus of Oslo City's healthcare department. 
\textcolor{black}{
The criticality of these applications requires rigorous and automated testing at various levels, such as integration and system testing. 
Conducting system and integration level testing is challenging as it requires creating a test infrastructure that considers the application as a whole. 
This infrastructure must incorporate various types of medical devices and third-party applications connected with an IoT-based healthcare application---a system under test (SUT). 
One of Oslo City's healthcare department's primary requirements is the development of a test infrastructure to enable automated and cost-effective testing of IoT-based healthcare applications. 
Thus, our innovation project aims to develop a test infrastructure that supports automated test execution at both the system and integration levels. 
In this regard, our experimental findings indicated that involving physical devices in the testing process could potentially result in device damage or the possibility of service blockages from the device vendors~\cite{sartaj2023testing}. 
Moreover, physically connecting multiple types of continually evolving medical devices is financially expensive, involves manual effort, and is infeasible in practice. 
Given these challenges, we proposed to employ DTs of medical devices into the test infrastructure~\cite{sartaj2023hita}. 
Our experiments revealed that substituting DTs for medical devices is a cost-effective and practical method to enable automated test execution~\cite{sartaj2023hita,sartaj2024modelbased}. 
The main role of DTs is to support automated testing of IoT-based healthcare applications at both system and integration levels, thereby alleviating the challenges associated with the involvement of physical devices in testing. 
}

A major challenge in creating high-fidelity DTs is simulating the non-deterministic environment of medical devices. 
The operating environment consists of various uncertain events that are necessary for a DT to reflect a medical device's behavior precisely and for calibration and synchronization of a DT. 
\textcolor{black}{
Another challenge in our context is determining the probability distribution that best represents the environmental uncertainties. 
This is because different vendors supply various medical device types, and the precise probability distribution each device follows is typically undisclosed to practitioners or is unknown. 
Therefore, determining the suitable probability distribution would facilitate the simulation of uncertain environmental scenarios that are necessary for testing (examples in \Cref{sec:motivation}).
}
Furthermore, rigorous testing of IoT-based healthcare applications necessitates the consideration of the environmental factors associated with medical devices~\cite{sartaj2024digital}. 
\textcolor{black}{
Various environmental factors, such as users, hardware components, and networks, can give rise to numerous uncertain scenarios. 
For instance, failure in network connection could disrupt the device's connection with the IoT application. 
}
In the next section, we elaborate on specific scenarios to provide a more detailed understanding of the problem.

\begin{figure}[htbp]
\centerline{\includegraphics[width=\linewidth, keepaspectratio]{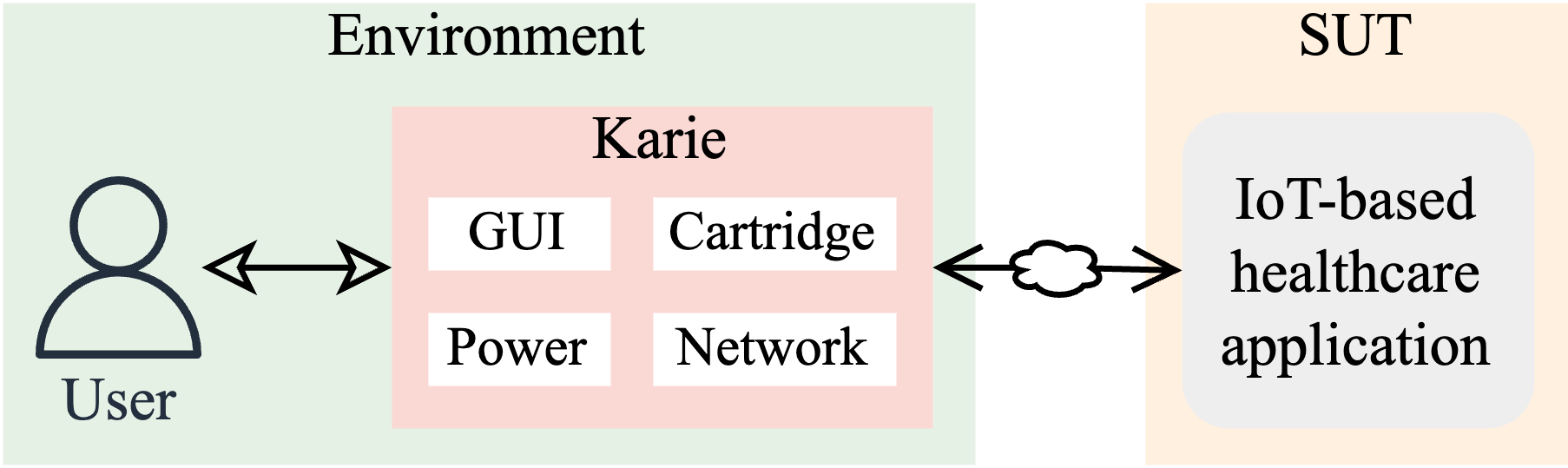}}
\caption{An example of an environment for the Karie medicine dispenser.}
\label{fig:ex-env}
\end{figure}

\subsection{Motivating Scenarios}\label{sec:motivation}
We illustrate the environmental cases of medical devices using an example of a Karie~\cite{karie} medicine dispenser integrated with our IoT-based healthcare application (SUT), as shown in \Cref{fig:ex-env}. 
Karie is an automatic medicine dispenser primarily designed to ensure the timely delivery of medications. 
It has an easy-to-use touch-driven graphical user interface (GUI) for various users, allowing seamless customization of device settings (e.g., language, alarm volume, and brightness). 
The types of users include patients, caretakers, or pharmacists. 
In addition, Karie contains a cartridge for medicine rolls, a rechargeable battery as a power source, and network cards for communication via 4G and WiFi. 
It also includes additional components like a camera, sensors, and medicine tray, although not depicted in \Cref{fig:ex-env} for simplicity.

Karie commences its operation by scanning the medication roll in a cartridge to ensure the correct medicine is available for the designated patient.
Following this, it automatically fetches medication plans, as prescribed by medical professionals, from an IoT-based healthcare application. 
Subsequently, it adheres to the medication plan, awaits the designated medication time, dispenses medicines at the specified hour, and triggers an alarm to remind the patient about their medication.
Before retrieving medicines from the tray, the patient presses a button to confirm, while simultaneously undergoing facial recognition through the integrated camera.
Whether a patient takes medicine or not, Karie sends notifications to the IoT-based healthcare application. 
Furthermore, it consistently updates the device's status, including power failures, network issues, hardware malfunctions, and other pertinent information.

Karie's operating environment has many uncertain scenarios related to user interactions, device hardware, and communication networks. 
Such scenarios are crucial for Karie's DT to calibrate and synchronize with the physical Karie and test the SUT. 
Some common scenarios, drawn from our experiences, are outlined below.

\begin{itemize}
    \item \textbf{Scenario 1 -- Network.} A patient misses medicine doses, and Karie fails to send notifications due to the network connectivity problem. This leads to two consequences: (i) Karie's DT being unsynchronized with the physical Karie, and (ii) the notification feature of the SUT going untested.

    \item \textbf{Scenario 2 -- Hardware.} Karie shuts down due to a low battery or power supply failure shortly after dispatching a notification to SUT. This breaks Karie DT's communication link with physical Karie, resulting in DT calibration and synchronization issues.  
    
    \item \textbf{Scenario 3 -- User Interactions.} A patient takes an overdose by tapping the travel medicine icon on the GUI. 
    Karie promptly reports this occurrence to the SUT, which, recognizing the urgency of the situation, assigns a task to the appropriate personnel for immediate attention and resolution on an emergency basis. Simulating such user interactions for Karie's DT is imperative to test SUT's task assignment and resolution features. 
\end{itemize}

On top of that, various medical devices, each with distinct hardware and network characteristics, combined with diverse user interactions involving patients, caretakers, and pharmacists, lead to numerous uncertain environmental scenarios. 
These scenarios are particularly pertinent to healthcare authorities (Oslo City, in our case) when testing IoT-based healthcare applications connected to DTs.
Motivated by this, we devise an approach in this paper for modeling and simulating the uncertain environment of medical devices DTs.

\section{\textcolor{black}{Approach Overview}}\label{sec:approach}

\begin{figure*}[htbp]
\centerline{\includegraphics[width=\textwidth, keepaspectratio]{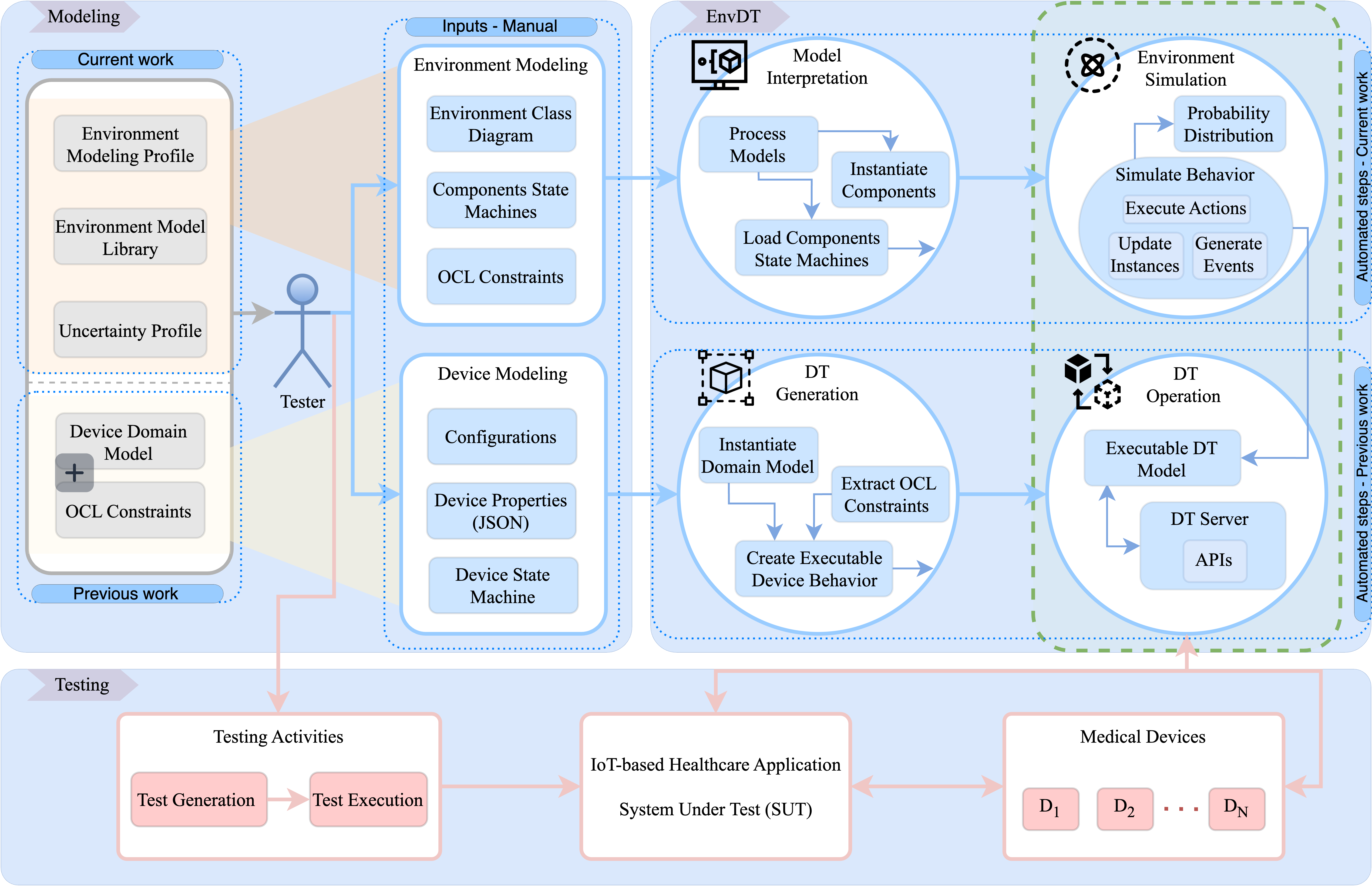}}
\caption{An overview of the \approach{}, including environment and device modeling steps, environment simulation from models, DT generation and operation, and the integration with an IoT-based healthcare application during testing.}
\label{fig:app}
\end{figure*}

\Cref{fig:app} presents an overview of the proposed approach (\approach{}).
Within our modeling methodology, we introduce a Unified Modeling Language (UML) profile specifically designed for modeling the environment of medical devices.
We also provide a UML model library, which serves as a comprehensive resource for modeling environment-related events. 
For modeling uncertain aspects of the environment, we utilize UML uncertainty profile (UUP)~\cite{zhang2019uncertainty}.  
\textcolor{black}{
Using the environment modeling profile and model library, a tester must model structural and behavioral aspects of a device's environment. 
During the creation of the structural model in the form of a UML class diagram, a tester needs to: (i) specify the environment components, (ii) add signal receptions from the UML model library, and (iii) create associations between the environment and DT classes. 
}
\textcolor{black}{
When creating behavioral models in the form of UML state machines, a tester needs to: (i) apply stereotypes from the environment profile, (ii) specify OpaqueBehavior (such as \emph{Entry} and \emph{DoActivity}) for each state, (iii) add Python code inside the body of each OpaqueBehavior specifying the state's internal behavior, and (iv) create submachines states that have an associated sub-behavior corresponding to an environment component (such as Battery or Network). 
}
Once the structural and behavioral models have been created, a tester must employ UUP to model uncertainties within these models. 
\textcolor{black}{
In our approach, we utilize the \emph{Belief} profile from UUP, where a tester must specify the belief element and the associated degree of belief in the form of probabilities. 
For this purpose, testers may rely on their knowledge and confidence level about uncertainties in various model elements. 
}
Finally, a tester needs to specify constraints on the environment models. 
The constraints must be defined using the Object Constraint Language (OCL), a standard language for specifying constraints on UML models~\cite{sartaj2019search,sartaj2020cdst}. 
\textcolor{black}{All the environment models including uncertainties and constraints are required to be created manually by testers, which is a common practice in modeling~\cite{sartaj2021testing,sartaj2024automated}. }

Upon completion of the modeling phase, the \approach{} phase begins. 
\textcolor{black}{
This phase comprises all the automated steps, as highlighted in \Cref{fig:app}. 
}The initial step involves interpreting the environment models. 
This encompasses loading the environment's structural models (represented as UML class diagrams), instantiating these models, and loading the behavioral models (in the form of UML state machines) of the environment components. 
With the instance models and state machines of the environment, we create a simulation of each environment component. 
\textcolor{black}{
More precisely, we traverse the state machines to extract behavioral aspects such as states along with their opaque behaviors and submachines, transitions, events, and associated uncertainties. 
During the traversal of state machines, we simulate the behavior of states by executing code corresponding to each state's opaque behaviors and submachines. 
While executing state behaviors, we consider various probability distributions (as outlined in~\cite{zhang2019uncertainty}) to simulate the uncertainties of the environmental components. 
Throughout the simulation, instance models of environment components are constantly updated with the changes in property values while evaluating OCL constraints. 
}
\textcolor{black}{
For the DT part, we adhere to the previous approach~\cite{sartaj2024modelbased} for DTs generation, which includes instantiating the DTs domain model, loading OCL constraints, and creating executable DTs behavior. 
To operate DTs together with environment simulations, we incorporate uncertain events handling in executable DT models. 
The subsequent steps, including setting up the DT server and APIs, remain the same as in the previous approach. 
}

During the testing phase, DTs integrated with an IoT-based healthcare application start their operation. 
While operating a DT, it goes through various states during behavior execution, e.g., \emph{Dispensing} state in the case of medicine dispensers. 
At the same time, the environment simulator asynchronously generates and sends various signal events to the DT. 
The DT handles these events and changes its state or properties accordingly. 
In response to requests from an IoT-based healthcare application, the DT will provide responses based on its current state within the uncertain environment. 

\textcolor{black}{
Since \approach{} is primarily designed to simulate uncertain environments for DTs, it requires simultaneous operation with DTs. 
Thus, we adopt DT generation and operation steps from previous work~\cite{sartaj2024modelbased}, as highlighted in \Cref{fig:app}. 
In our approach, we introduce several new environment-related aspects. 
These include the environment modeling methodology with a UML profile and a model library, the model interpretation steps, and the uncertainty-wise environment simulation. 
Furthermore, given that DT generation and operation steps rely on the device models, we enhance the modeling aspect to generate and operate DTs alongside their corresponding environment simulations. 
The specific enhancements related to modeling include (i) annotating structural and behavioral models of a device with environment and uncertainty profiles, (ii) associating signal receptions (from the model library) in structural models, (iii) interconnecting the device's structural parts with their environmental components, and (iv) specifying associated environmental components behaviors in connection with device's behavioral model. 
Using the enhanced models, the DT generation process remains the same. 
During the operation of DTs, the executable DT models operate while handling uncertain environmental events. 
In the subsequent sections, we elaborate on these enhancements in detail. 
}

\section{\textcolor{black}{Environment Modeling Utilities}}\label{sec:modutils} 
In the next sections, we discuss our UML profile for modeling the environment, UML model library, and UML uncertainty profile (adopted from~\cite{zhang2019uncertainty}).

\begin{figure*}[htbp]
\centerline{\includegraphics[width=13cm, keepaspectratio]{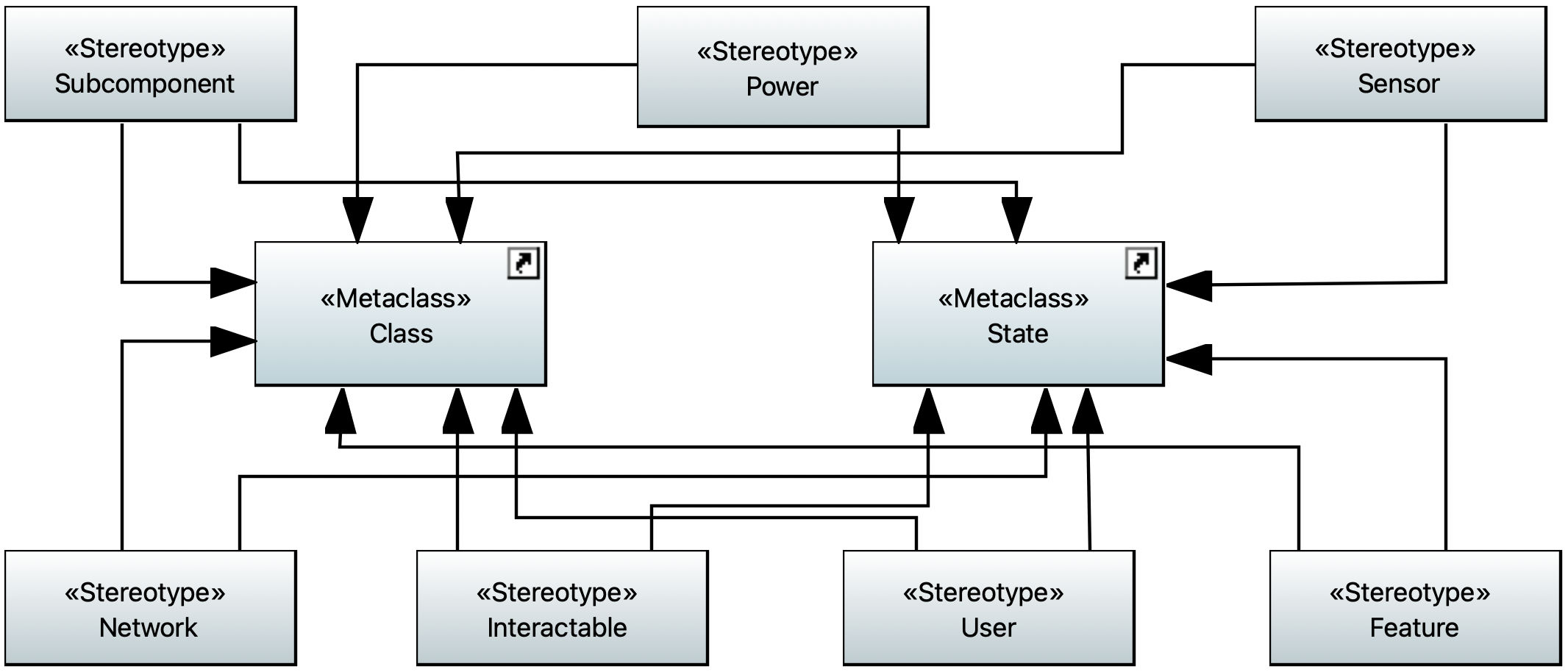}}
\caption{A UML profile for modeling the environment of medical devices.}
\label{fig:env-profile}
\end{figure*}

\begin{figure*}[htbp]
\centerline{\includegraphics[width=\textwidth, keepaspectratio]{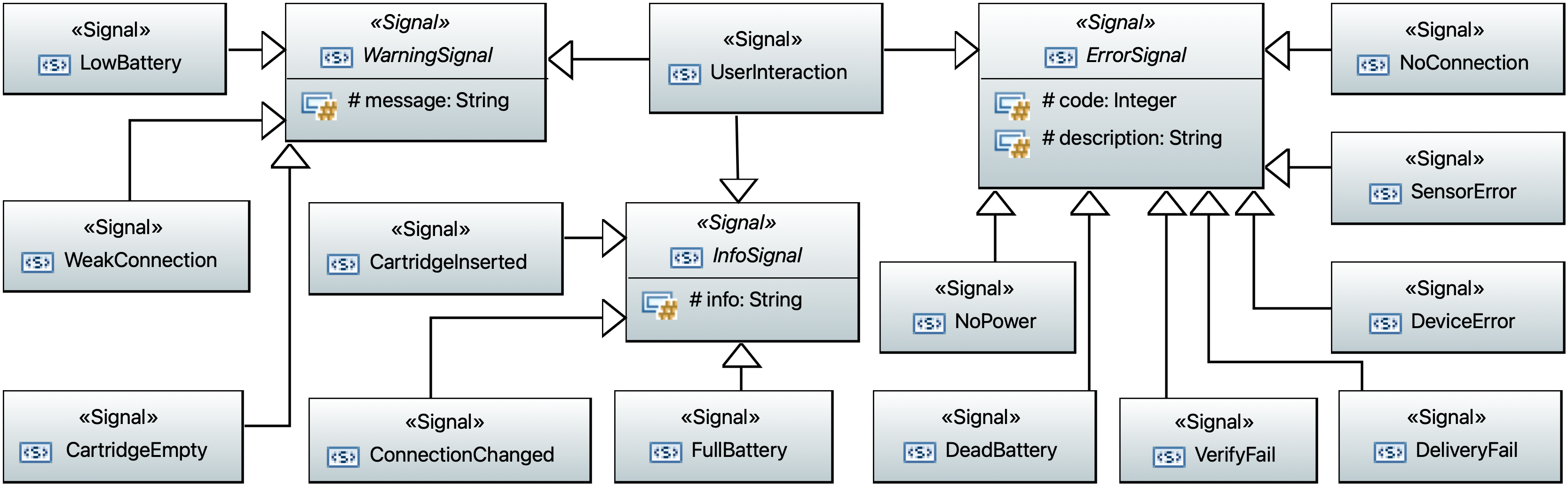}}
\caption{A UML model library for modeling signal events of environment.}
\label{fig:env-ml}
\end{figure*}

\subsection{UML Profile for Environment Modeling}\label{ssec:profile}
UML is a standard multi-purpose modeling language with a broad scope, making it adaptable to various domains. 
To apply UML in a particular domain, UML provides an extension mechanism known as profiles. 
Using the UML profiling feature, we develop a simplified extension specifically for modeling the environment of medical devices. 
This profile is designed to enable industry practitioners to effortlessly model the environment.

\Cref{fig:env-profile} shows our proposed UML profile for modeling the environment of medical devices. 
The profile consists of stereotypes; \guillemetleft Subcomponent\guillemetright, \guillemetleft Power\guillemetright, \guillemetleft Sensor\guillemetright, \guillemetleft Network\guillemetright, \guillemetleft Interactable\guillemetright, \guillemetleft User\guillemetright, and \guillemetleft Feature\guillemetright. 
The stereotype \guillemetleft Subcomponent\guillemetright~represents detachable parts (e.g., cartridge) of a medical device. 
The stereotype \guillemetleft Power\guillemetright~symbolizes power sources (e.g., rechargeable battery) of a medical device. 
The stereotype \guillemetleft Sensor\guillemetright~describes sensors attached to a medical device, e.g., a motion sensor with an activity monitoring device. 
The stereotype \guillemetleft Network\guillemetright~represents communication sources of a medical device. 
The stereotype \guillemetleft Interactable\guillemetright~symbolizes UI elements of a medical device with which a user (patient) operates the device. 
The stereotype \guillemetleft User\guillemetright~denotes potential medical device users, typically including a patient, caretaker, or medical specialist. 
The stereotype \guillemetleft Feature\guillemetright~characterizes the core functionalities of a medical device. 
All profile stereotypes extend from UML \emph{Class} and \emph{State} meta-classes because environmental components can have both structural and behavioral aspects. 
This is important for determining the relationships between behaviors and classes associated with environmental components, as well as for monitoring instances that change in response to behavioral executions during environment simulation.

\subsection{UML Model Library for Environment Events}\label{ssec:ml}
We developed a UML model library (shown in \Cref{fig:env-ml}), which comprises common signal events that we encountered during our experience with different medical devices.
We have classified concrete signal events into three categories---\emph{InfoSignal}, \emph{WarningSignal}, and \emph{ErrorSignal}---each modeled as abstract signals. 
The \emph{InfoSignal} category represents basic information-related signal events, which comprise \emph{CartidgeInserted}, \emph{ConnectionChanged}, and \emph{FullBattery}. 
The \emph{WarningSignal} category describes warning notifications containing signals for \emph{LowBattery}, \emph{WeakConnection}, and \emph{CartidgeEmpty}. 
Similarly, the \emph{ErrorSignal} category defines error messages or critical notifications which consist of \emph{NoPower}, \emph{DeadBattery}, \emph{VerifyFail}, \emph{DeliveryFail}, \emph{DeviceError}, \emph{SensorError}, and \emph{NoConnection} signals. 
Signal events associated with \emph{UserInteraction} can be categorized under info, warning, or error signals.
All the above-described signal events need to be incorporated when modeling signal receptions within the structural aspect of the environment components.
Additionally, these events must be associated with state transitions (precisely with triggers) within state machines that represent behavioral aspects of environment components.

\subsection{UML Uncertainty Profile}\label{ssec:uup}

To specify environmental uncertainties in the models, we adopt the UML uncertainty profile (UUP)~\cite{zhang2019uncertainty}. 
\textcolor{black}{
The rationale for choosing UUP is that it provides a comprehensive modeling methodology to capture and model environmental uncertainties for testing CPS, which aligns well with our approach. 
}
UUP consists of UML profiles named \emph{Belief} profile, \emph{Measurement} profile, and \emph{Uncertainty} profile. 
Moreover, it includes model libraries called \emph{Vagueness}, \emph{Ambiguity}, and \emph{Probability} to define various uncertainty measures. 
As the UUP work provides detailed guidelines for modeling uncertainties, we describe its implementation within our context. 

In our approach, it is necessary to specify a tester's beliefs and uncertainties within both structural and behavioral models. 
The measurement-related uncertainties need to be specified only in structural models. 
For the uncertainty measure, it is required to leverage the \emph{Probability} model library. 
Utilizing this specific library allows for a more accurate and comprehensive representation of uncertainty within the models. 
In addition, this library encompasses various probability distributions, which are essential for simulating environmental uncertainties in our case. 
These probability distributions provide the probabilities of occurrence of different outcomes in a process, thereby allowing for more uncertain simulations of the environment.

\begin{figure*}[htbp]
\centerline{\includegraphics[width=\textwidth, keepaspectratio]{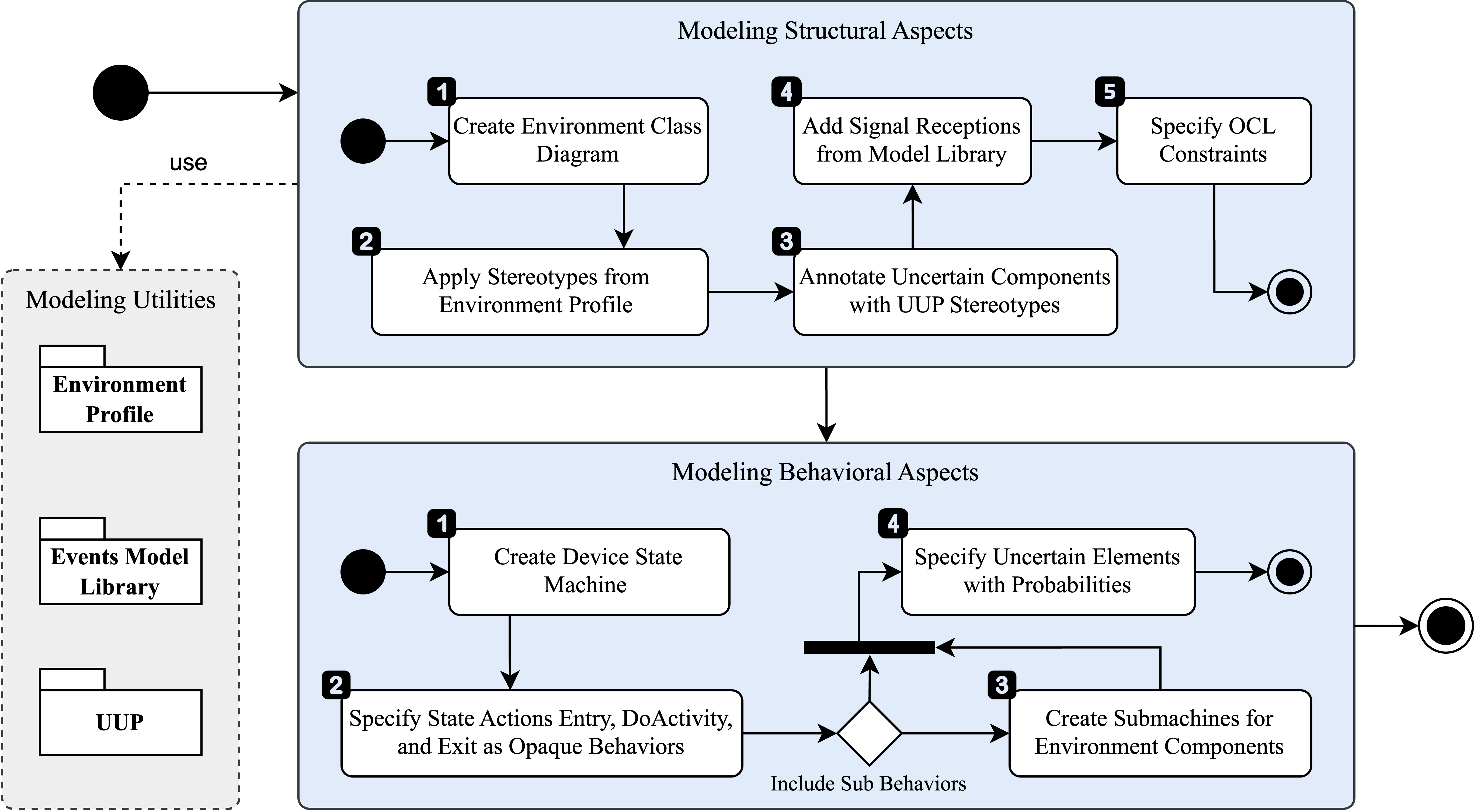}}
\caption{Workflow of the methodology for modeling structural and behavioral aspects of a device's environment.}
\label{fig:methodology}
\end{figure*}

\section{\textcolor{black}{Methodology to Model and Simulate Environment}}\label{sec:envmodels}
\Cref{fig:methodology} presents an overview of the methodology for modeling a device's environment. 
With the environment modeling utilities (discussed in \Cref{sec:modutils}), we first model the structural aspects of the environment. 
This includes creating an environment class diagram, applying stereotypes from the environment profile, annotating uncertain environmental components using UUP, adding signal receptions from the signal events model library, and specifying OCL constraints. 
We elaborate on these steps in detail in \Cref{sec:structmodel} using a medicine dispenser (Karie~\cite{karie}) as an illustrative example. 
Once the environment class diagram is developed, we proceed to model the device's behavior alongside its environmental components.  
During behavioral modeling, we create a device state machine as an owned behavior of the device's class. 
In each state of the device state machine, we specify OpaqueBehavior in the form of \emph{Entry}, \emph{DoActivity}, and \emph{Exit}. 
If a state requires the internal behavior of an environment component, we create a submachine to model the component's behavior. 
Subsequently, the final step involves modeling the uncertain behavioral elements and their respective probabilities within all state machines. 
In \Cref{sec:devbehmodel,sec:compbehmodel}, we elaborate on these steps using the Karie medicine dispenser example.

\subsection{Modeling Structural Aspects}\label{sec:structmodel}
\begin{figure*}[htbp]
\centerline{\includegraphics[width=\textwidth, keepaspectratio]{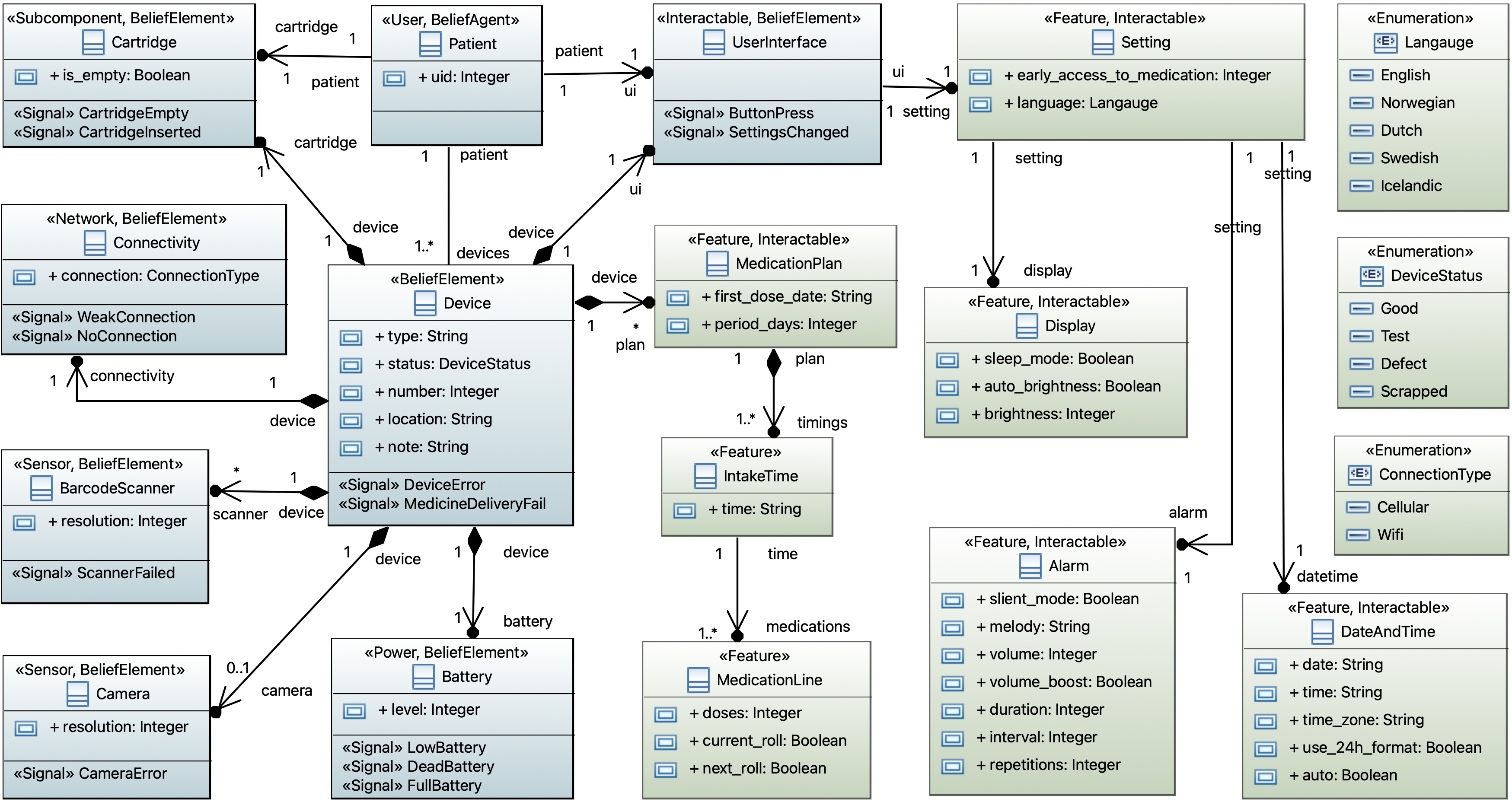}}
\caption{A UML class diagram of a medicine dispenser (Karie) visualizing associations between the environment and the DT. Classes depicted in blue signify the components of the environment, whereas those represented in green correspond to DT concepts.}
\label{fig:env-dt-cd}
\end{figure*}

Our approach requires modeling structural aspects of the environment in the form of a UML class diagram, which must establish relationships with the DT structural model. 
\Cref{fig:env-dt-cd} shows a UML class diagram for the Karie medicine dispenser, consisting of classes for the environment components (depicted in blue) and DT concepts (highlighted in green). 
The DT classes are adopted from the previous model-based DT generation approach~\cite{sartaj2024modelbased}. 
The environmental components represent classes for \emph{Patient}, \emph{Device}, \emph{UserInterface}, \emph{Cartridge}, \emph{Connectivity}, \emph{BarcodeScanner}, \emph{Camera}, and \emph{Battery}. 
A patient can have one or more dispensers (modeled as \emph{Device}). 
A \emph{Device} is composed of a \emph{UserInterface} for user interactions, \emph{Cartridge} for medicine rolls, \emph{Connectivity} for network connection, \emph{BarcodeScanner} for verification of medicine roll, \emph{Camera} for facial recognition, and \emph{Battery} for power supply. 
All the environmental classes are stereotyped using environment profile and UUP.
For example, a patient is a \emph{User} of a medicine dispenser and is a \emph{BeliefAgent} with a certain degree of belief about the device uncertainties. 
In addition to profile stereotypes, signal events are modeled as receptions in various components' classes. 
For instance, the \emph{Connectivity} class includes a warning signal (\emph{WeakConnection}) and an error signal (\emph{NoConnection}).

\paragraph{Specifying Constraints}
The concepts and properties modeled in the environment class diagram can have constraints on their instances and values. 
Such constraints need to be specified on the class diagram using OCL. 
While instantiating classes of environment components, OCL constraints are essential to ensure the correct configurations of the instance models, e.g., number of instances and property values. 
For the environment class diagram shown in \Cref{fig:env-dt-cd}, \Cref{lst:env-constraints} presents an excerpt of OCL constraints specified on various environment components. 
Constraint \emph{C1} defines that a device's battery level must be from 0 to 100\%. 
Constraint \emph{C2} specifies that a patient cannot have two devices with the same serial numbers. 
Constraint \emph{C3} restricts a patient's ID (\emph{uid}) to be a positive non-zero number. 
Similarly, constraint \emph{C4} ensures that a patient can have at least one device and at most \textit{N} devices, where \textit{N} is specified by healthcare authorities (e.g., Oslo City healthcare department). 

\begin{lstlisting}[label=lst:env-constraints, language=ocl, caption={A snippet of OCL constraints on various environment model elements.}, linewidth=7.5cm, numbers=none]
-- Values range for the device's battery level 
C1: context Device inv: self.battery.level >= 0 and self.battery.level <= 100
-- A patient cannot have two same devices 
C2: context Patient inv: self.devices->forAll(d1, d2 | d1.number <> d2.number)
-- Patients' ID cannot be a negative or zero
C3: context Patient inv: self.uid > 0
-- The number of devices a patient can own
C4: context Patient inv: self.devices->size() > 0 and self.devices->size() <= N
\end{lstlisting}

\begin{figure*}[htbp]
\centerline{\includegraphics[width=\textwidth, keepaspectratio]{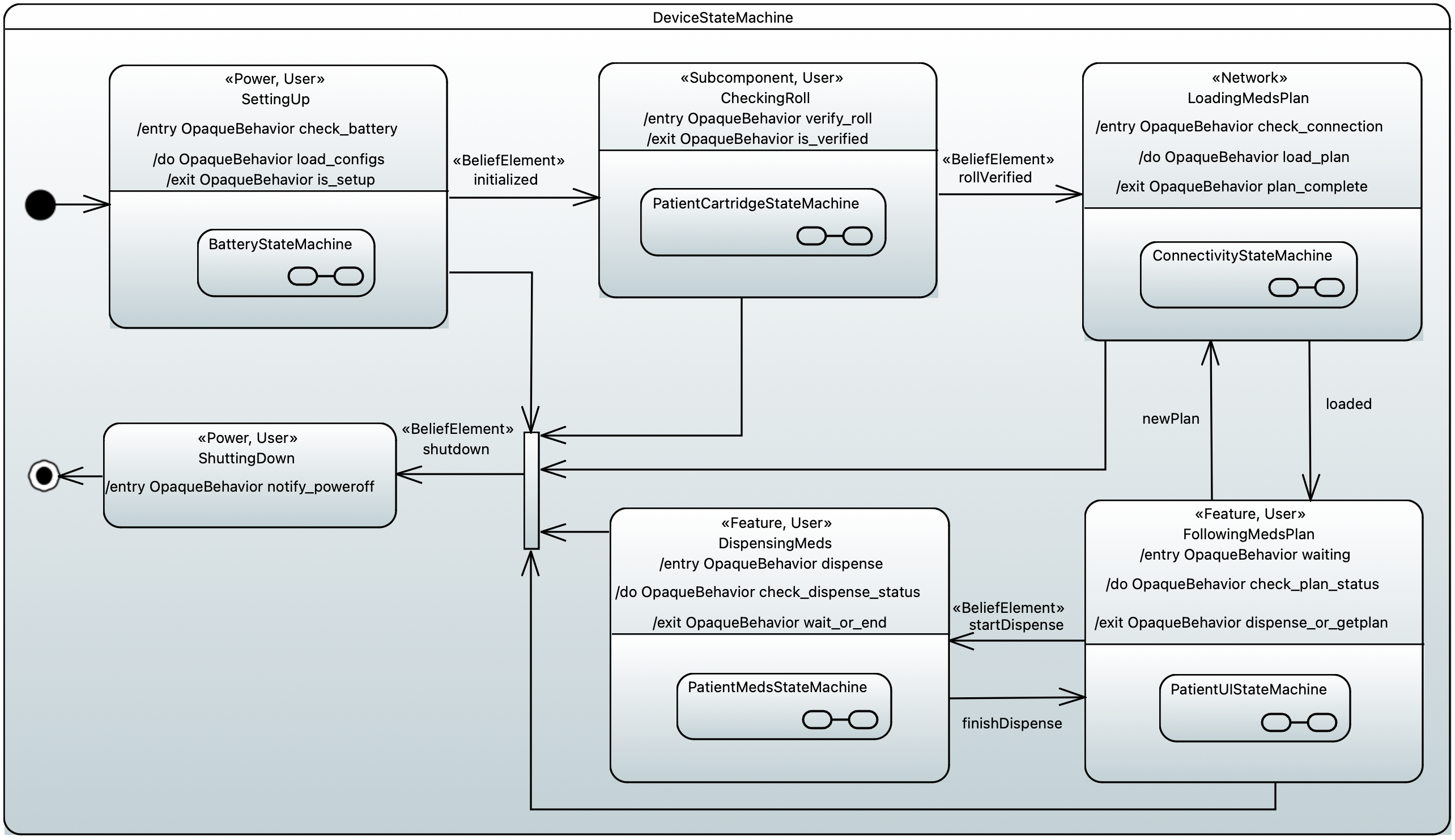}}
\caption{A UML state machine representing medicine dispenser's behavior--an extended version of the one presented previously~\cite{sartaj2024modelbased}. \textcolor{black}{The composite states indicate the internal behavior of states, modeled as submachines. The associated submachines are fully illustrated in \Cref{fig:battery-sm,fig:patient-crtg-sm,fig:conn-sm,fig:patient-ui-sm,fig:patient-meds-sm}: \Cref{fig:patient-meds-sm} for \textit{PatientMedsStateMachine}, \Cref{fig:patient-ui-sm} for \textit{PatientUIStateMachine}, \Cref{fig:patient-crtg-sm} for \textit{PatientCartridgeStateMachine},  \Cref{fig:conn-sm} for \textit{ConnectivityStateMachine}, and \Cref{fig:battery-sm} for \textit{BatteryStateMachine}.}}
\label{fig:dev-dt-sm}
\end{figure*}

\subsection{Modeling Device Behavioral Aspects}\label{sec:devbehmodel}
After the structural modeling of the environment and DT, the subsequent step involves modeling the device's behavior utilizing UML state machines. 
These state machines should be created as owned behaviors of the device class and must contain the states, transitions, and events associated with the device---aligning with the device's DT behavior~\cite{sartaj2024modelbased}.  
Moreover, to model the device's behavior, our approach requires using the environment profile, UUP, and signal events model library. 

\Cref{fig:dev-dt-sm} shows a UML state machine representing a device's behavior for the Karie medicine dispenser example. 
The device state machine contains states named \emph{SettingUp}, \emph{CheckingRoll}, \emph{LoadingMedsPlan}, \emph{FollowingMedsPlan}, \emph{DispensingMeds}, and \emph{ShuttingDown}. 
When a user turns on the device, it starts with \emph{SettingUp} state, which is annotated with \guillemetleft Power\guillemetright~and \guillemetleft User\guillemetright~stereotypes from the environment profile, indicating the user's involvement and power source requirement. 
During \emph{SettingUp} state, the first action is to check the battery level---modeled as an \emph{Entry} action. 
This process requires invoking the \emph{Battery} state machine, which is included as a \emph{Submachine}. 
If the battery level is sufficient for device operation, it begins to load device configurations such as alarm settings---modeled as an \emph{DoActivity} action. 

Once the device's initialization is finished (i.e., \emph{Exit} action is executed), the next state is \emph{CheckingRoll}. 
The transition from \emph{SettingUp} to \emph{CheckingRoll} state is annotated with \guillemetleft BeliefElement\guillemetright~stereotype from UUP. 
\emph{CheckingRoll} state involves verification of the medicine roll inside the cartridge. 
After successful verification, the next state is \emph{LoadingMedsPlan}, in which the device establishes a network connection to load the medication plan from an IoT-based healthcare application. 
Once the medication plan is loaded, the next state is \emph{FollowingMedsPlan}. 
In this state, the device remains in a waiting mode, continuously checking for the scheduled time of the upcoming medicine dose. 
At the dose intake time, it transitions to \emph{DispensingMeds} state, which initiates the dispensing process. 
When dispensing finishes, it transitions back to \emph{FollowingMedsPlan} and waits for the next dose time. 
During each state, the device may end up in \emph{ShuttingDown} state due to low battery, failure, or button press by a user. 
Therefore, \guillemetleft BeliefElement\guillemetright~stereotype is applied to these transitions. 

When modeling device state machines, three factors require careful consideration. 
First, stereotypes from the environment profile must be applied to each device state to signify an association with the necessary environment component. 
Secondly, the tasks to be executed in a specific state must be modeled \emph{Entry}, \emph{DoActivity}, or \emph{Exit} actions using \emph{OpaqueBehavior} to define precise actions. 
Additionally, if a state requires the internal behavior of a device component (e.g., \emph{Connectivity} state machine inside \emph{LoadingMedsPlan}), the \emph{Submachine} for that particular component must be specified.
Third, transitions modeled with the \guillemetleft BeliefElement\guillemetright~stereotype should encompass the corresponding probabilities, which can be determined based on the tester's experience. 
\textcolor{black}{
Specifically, for each element modeled with the \guillemetleft BeliefElement\guillemetright~stereotype, a tester must specify a \emph{beliefDegree} of the \emph{Measurement} type. 
This \emph{Measurement} includes \emph{measure} property, as defined in UUP. 
Each \emph{measure} has a textual \emph{description} and a \emph{kind} of \emph{MeasureKind} type. 
Among UUP-provided measure kinds, a tester needs to select \emph{Probability}. 
While various probability scales (e.g., \emph{Probability\_7Scale}) were utilized in the UUP work~\cite{zhang2019uncertainty}, our approach requires a precise probability value in the range [0, 1]. 
These probability values reflect a tester's belief about a particular model element. 
For example, if a tester specifies a probability value of 0.9 for the \textit{shutdown} as shown in \Cref{fig:dev-dt-sm}, it indicates the tester's confidence level of 90\% that a device will transition into a \emph{ShuttingDown} state in the event of critical battery or power failure. 
The probability values associated with belief elements, when specified in this way, are used in our approach during simulations. 
It is important to note that a tester must adhere to the same method for modeling uncertainty on each model element with the \guillemetleft BeliefElement\guillemetright~stereotype. 
Furthermore, a tester must utilize UUP's \emph{Probability} model library to specify the probability distribution with the necessary properties. 
For instance, if \emph{UniformDistribution} is selected, the corresponding properties, i.e., \emph{min} and \emph{max} must be specified. 
}

\begin{figure*}[htbp]
\centerline{\includegraphics[width=13cm, keepaspectratio]{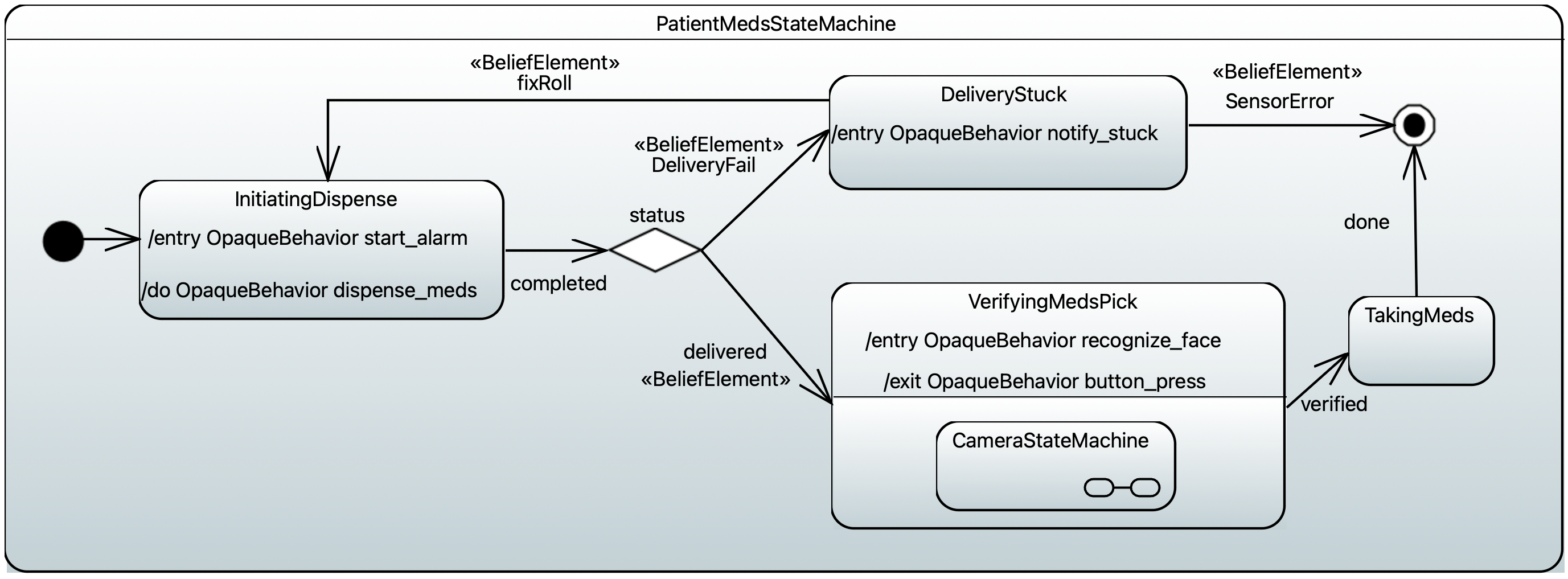}}
\caption{A UML state machine representing a patient's medicine intake process. \textcolor{black}{The internal behavior of the \textit{VerifyingMedsPick} state, modeled as a submachine, is fully depicted in \Cref{fig:camera-sm} for \textit{CameraStateMachine}.}}
\label{fig:patient-meds-sm}
\end{figure*}

\subsection{Modeling Environment Components Behavior}\label{sec:compbehmodel}
After modeling the device's behavior, the subsequent step is to model the behavior of various environment components using modeling utilities presented in \Cref{sec:modutils}.   
This includes modeling the behavior (in the form of UML state machines) of the user(s) as well as the components associated with the device.
These state machines should be created as owned behaviors within the class of the corresponding component.
For the ongoing example of a medicine dispenser, a patient as a user can have behaviors for medicine intake, cartridge fillings, and UI interaction with the device. 
Moreover, behavior-exhibiting device components for the medicine dispenser example are \emph{Connectivity}, \emph{Battery}, \emph{Scanner}, and \emph{Camera}. 
In the following, we explain the behavior of each environment component. 

\begin{figure*}[htbp]
\centerline{\includegraphics[width=10cm, keepaspectratio]{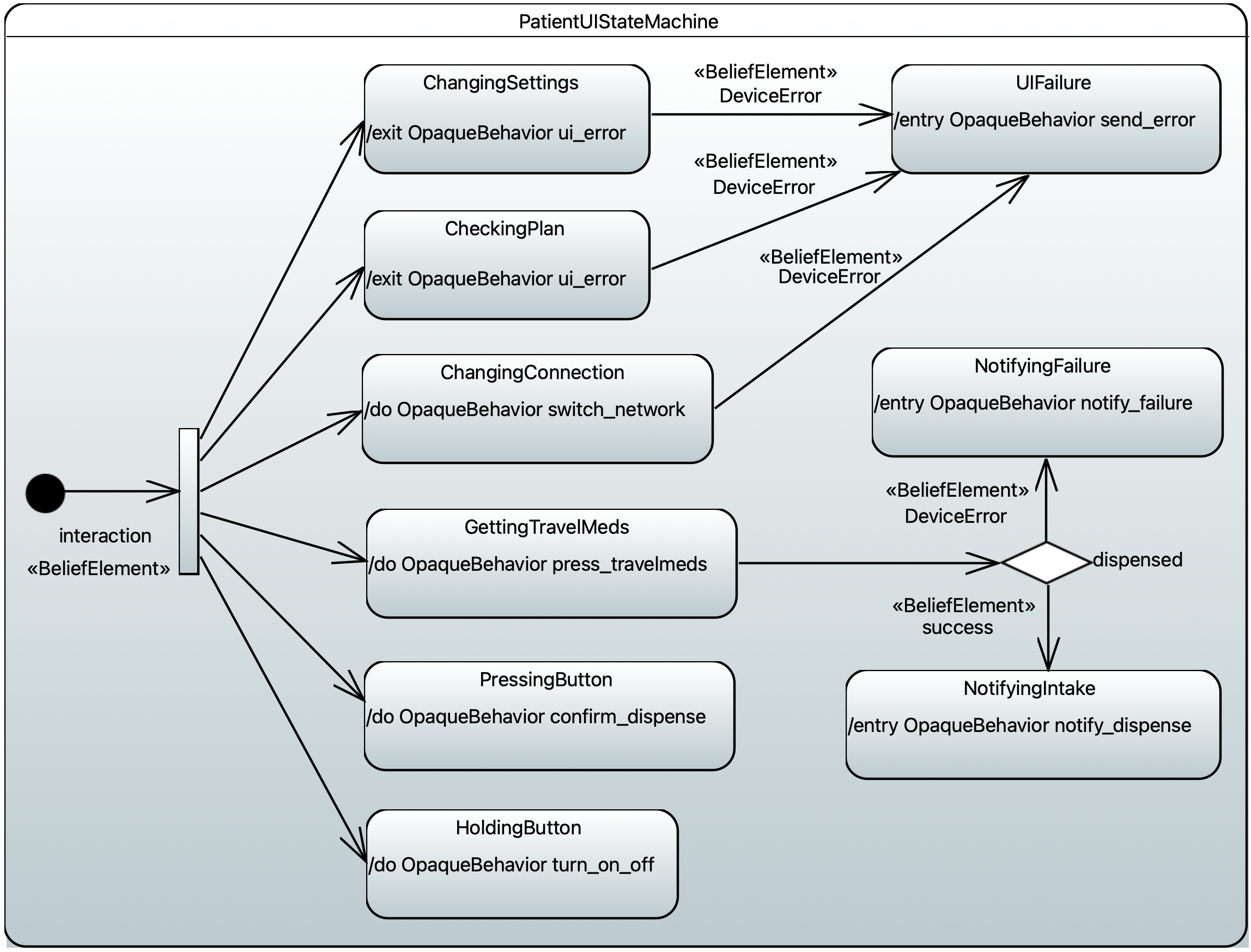}}
\caption{A UML state machine illustrating a patient's  interactions with the device UI.}
\label{fig:patient-ui-sm}
\end{figure*}

\begin{figure*}[htbp]
\centerline{\includegraphics[width=10cm, keepaspectratio]{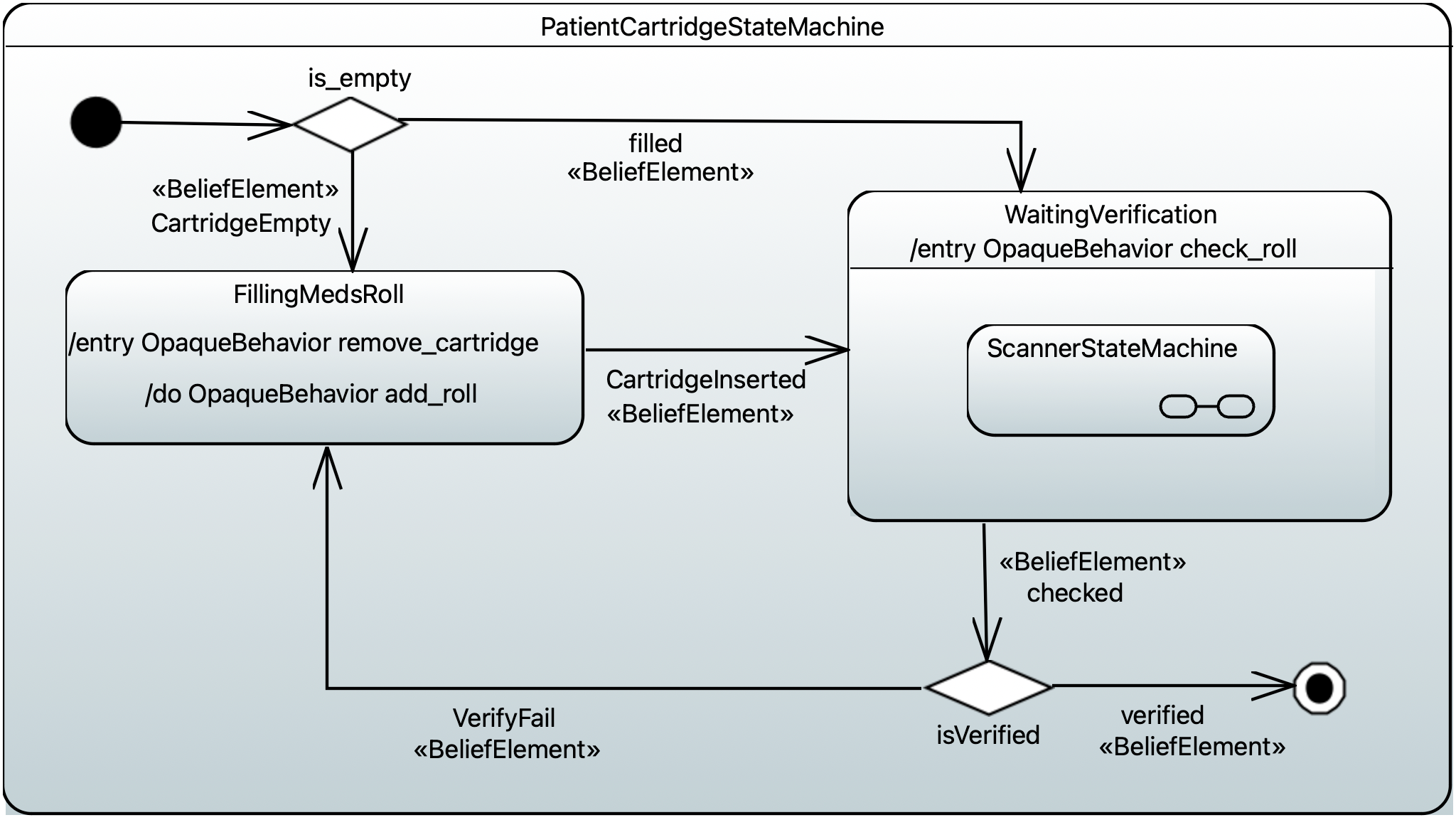}}
\caption{A UML state machine expressing a patient's cartridge filling procedure. \textcolor{black}{ The internal behavior of the \textit{WaitingVerfication} state, modeled as a submachine, is fully depicted in \Cref{fig:scanner-sm} for \textit{ScannerStateMachine}.}}
\label{fig:patient-crtg-sm}
\end{figure*}

\begin{figure*}[htbp]
\centerline{\includegraphics[width=11cm, keepaspectratio]{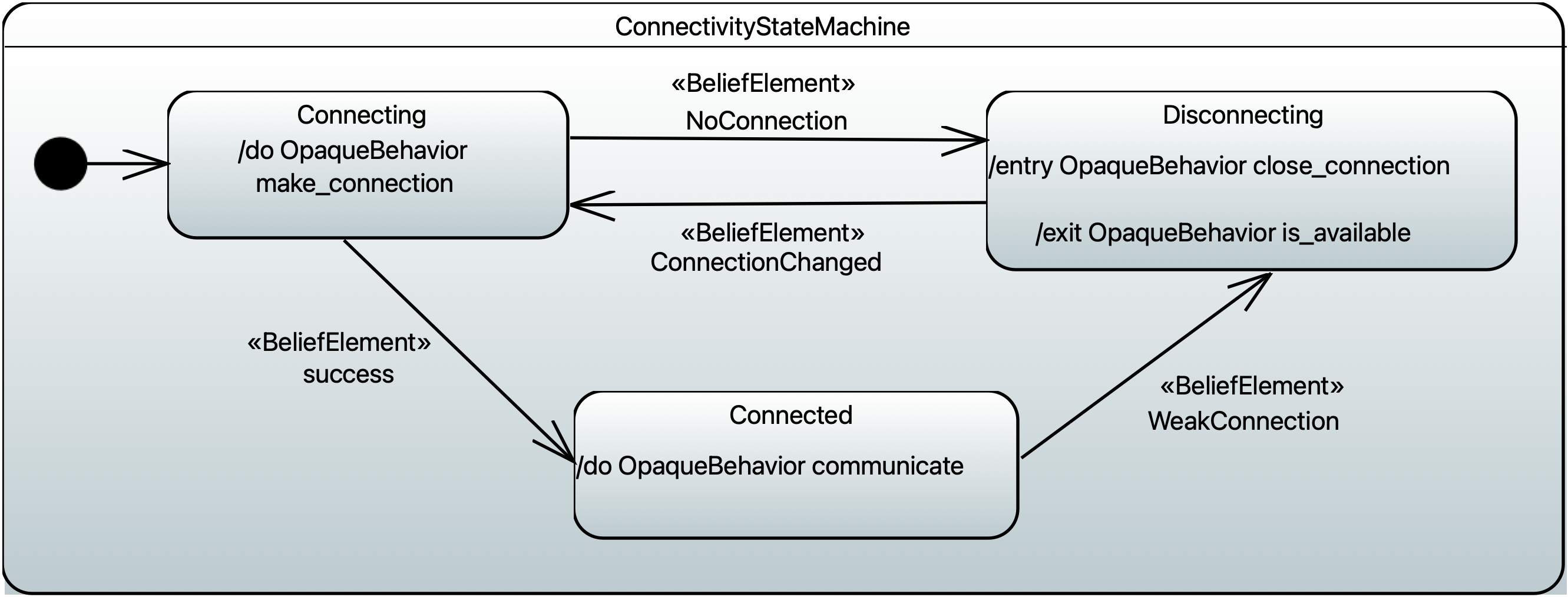}}
\caption{A UML state machine representing a device's network connectivity process.}
\label{fig:conn-sm}
\end{figure*}

\begin{figure}[htbp]
\centerline{\includegraphics[width=7.8cm, height=8cm, keepaspectratio]{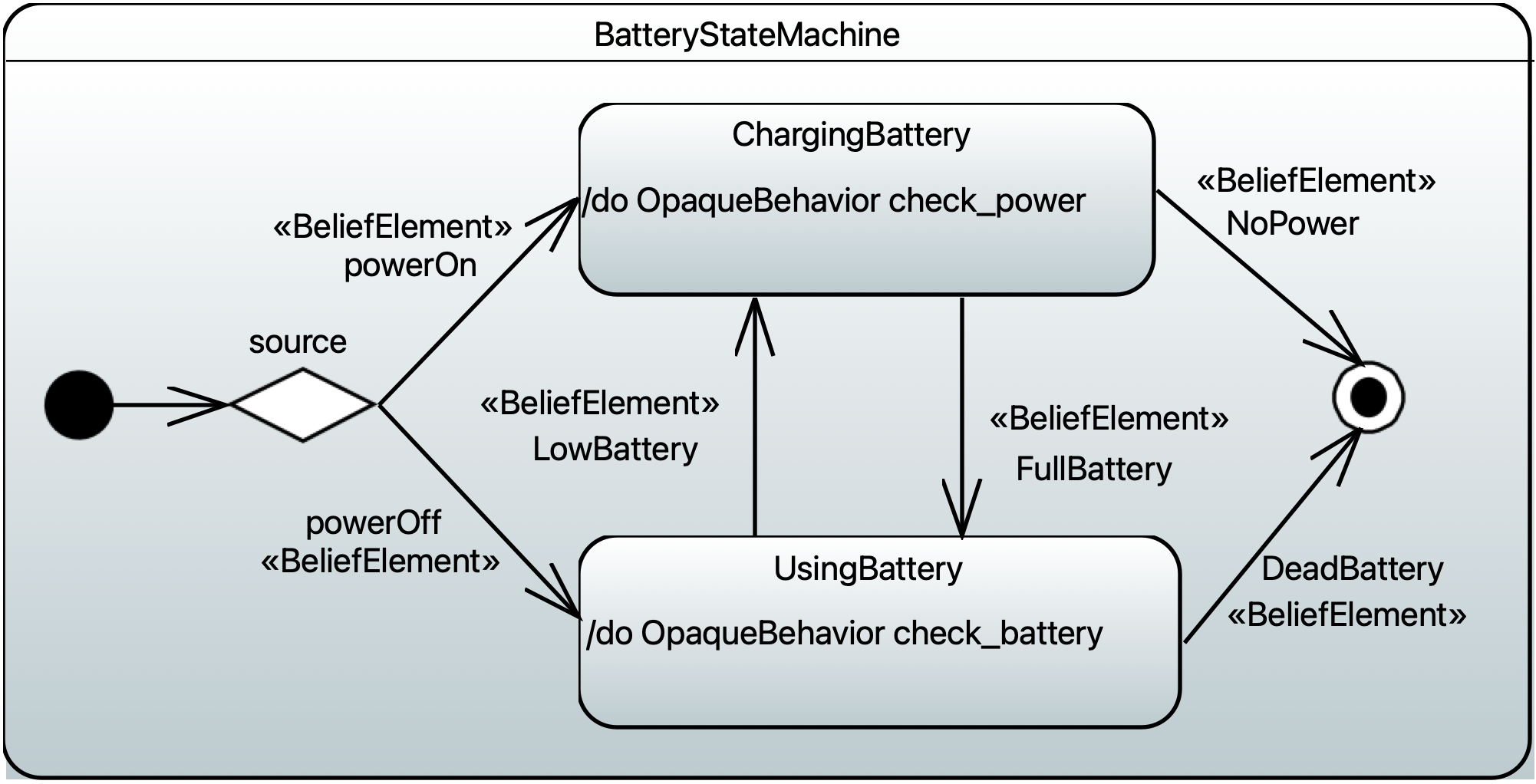}}
\caption{A UML state machine illustrating a device's battery and its charging mechanism.}
\label{fig:battery-sm}
\end{figure}

\subsubsection{Device User Behavior}
The user behavior modeled as state machines must capture different types of interactions a user might have with a specific device. 
Continuing with the example of the Karie medicine dispenser, \Cref{fig:patient-meds-sm,fig:patient-ui-sm,fig:patient-crtg-sm} depict three types of common behavioral interactions of a patient with dispensers, i.e., medication intake, UI interactions, and cartridge filling process. 
Each user behavior is modeled as UML state machines using UUP, and model library for signal events. 

The patient's medication behavior (as shown in \Cref{fig:patient-meds-sm}) starts with \emph{InitiatingDispense} state. 
At the specified medication time, the dispenser rings an alarm, which begins the medicine dispensing process. 
This transition is modeled with \guillemetleft BeliefElement\guillemetright~stereotype because it may lead to two situations: (i) medicine is delivered successfully, or (ii) the medicine pack is stuck. 
In the case of successful medicine delivery, the resultant state is \emph{VerifyingMedsPick}, in which the patient is recognized with a camera after pressing the confirmation button.  
After verification, the next state is \emph{TakingMeds} when the patient removes medicine from a pouch. 
When the medicine pack gets stuck, it transitions to the \emph{DeliveryStuck} state. 
This needs fixing the roll and restarting the dispensing process.  

The patient's behavior for UI interactions is illustrated in \Cref{fig:patient-ui-sm}. 
This includes different types of interaction states; \emph{ChangingSettings}, \emph{CheckingPlan}, \emph{ChangingConnection}, \emph{GettingTravelMeds}, \emph{PressingButton}, and \emph{HoldingButton}. 
The state \emph{ChangingSettings} refers to user interactions for modifying device settings, e.g., adjusting alarm volume or melody.  
The \emph{CheckingPlan} state represents user interactions intended to review medication plans. 
Similarly, \emph{ChangingConnection} state symbolizes the user's actions to modify the network connection, such as switching between WiFi or 4G connections.
The states \emph{ChangingSettings}, \emph{CheckingPlan}, and \emph{ChangingConnection} may lead to \emph{UIFailure}. 
The \emph{GettingTravelMeds} state can transition to either \emph{NotifyingIntake} state following successful dispensation, or the \emph{NotifyingFailure} in case of failure due to device error.
All transitions involving uncertainties are modeled with \guillemetleft BeliefElement\guillemetright~stereotype. 
The states \emph{PressingButton} and \emph{HoldingButton} represent scenarios where a user either presses a button to confirm medication intake or holds a button for a certain duration to turn off the device.

The state machine in \Cref{fig:patient-crtg-sm} demonstrates a patient's cartridge filling procedure. 
If the cartridge is empty, the device generates \emph{CartridgeEmpty} event, leading to the \emph{FillingMedsRoll} state. 
During this state, a patient removes the cartridge and inserts a medication roll. 
After filling the cartridge, the patient inserts it into the dispenser and waits for the verification (i.e., \emph{WaitingVerification} state) of the medicine roll. 
The verification of the medicine roll is performed by a barcode scanner, thus requiring \emph{Scanner} submachine. 
If the medicine roll is verified successfully, the process finishes. 
In the case of unverified medicine rolls, patients need to refill the cartridge with another medicine roll.

\subsubsection{Device Connectivity Behavior}
\Cref{fig:conn-sm} shows a UML state machine for a dispenser's connectivity behavior. 
This includes states \emph{Connecting}, \emph{Connected}, and \emph{Disconnecting}. 
The state machine initiates with the \emph{Connecting} state in which the device tries making a network connection. 
In the case of connection availability and good strength, it transitions to the \emph{Connected}. 
In the case of a weak network or no connection available, in either \emph{Connecting} or \emph{Connected} states, the next state becomes \emph{Disconnecting}. 
It remains in \emph{Disconnecting} state until the connection is restored or connection types are changed (e.g., switching from WiFi to 4G).
The transitions involving uncertainties in connectivity behavior are annotated with \guillemetleft BeliefElement\guillemetright~stereotype.

\begin{figure*}[htbp]
\centerline{\includegraphics[width=13cm, keepaspectratio]{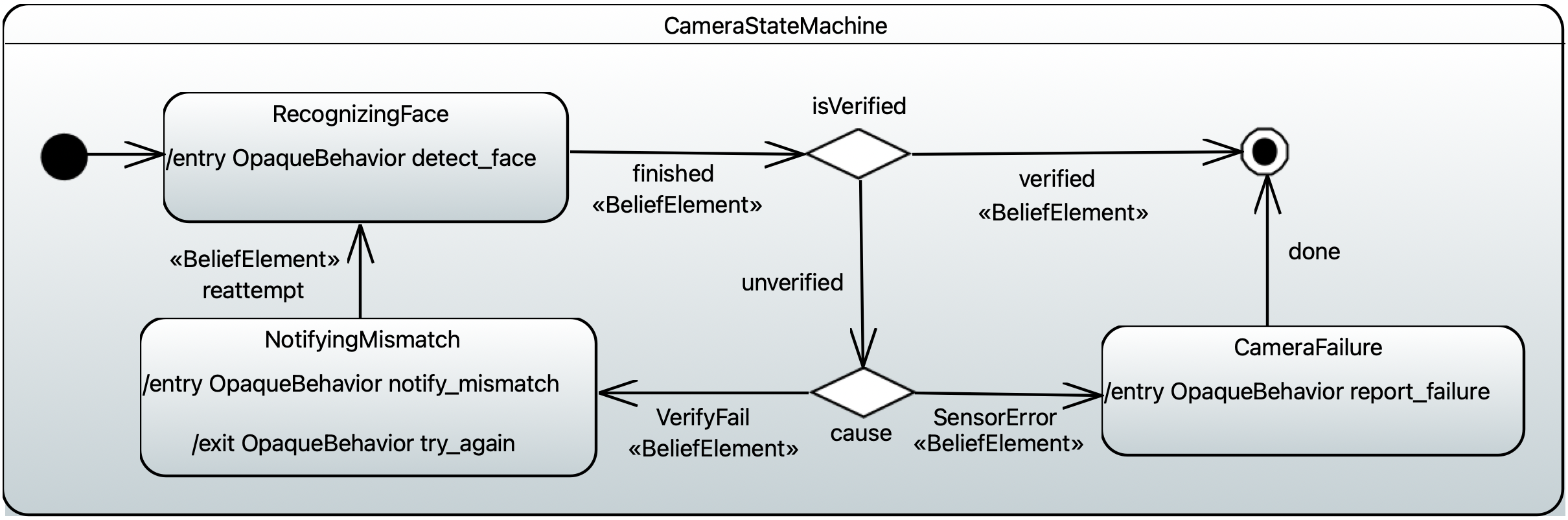}}
\caption{A UML state machine for a device's camera depicting camera facial recognition function.}
\label{fig:camera-sm}
\end{figure*}

\begin{figure*}[htbp]
\centerline{\includegraphics[width=13cm, keepaspectratio]{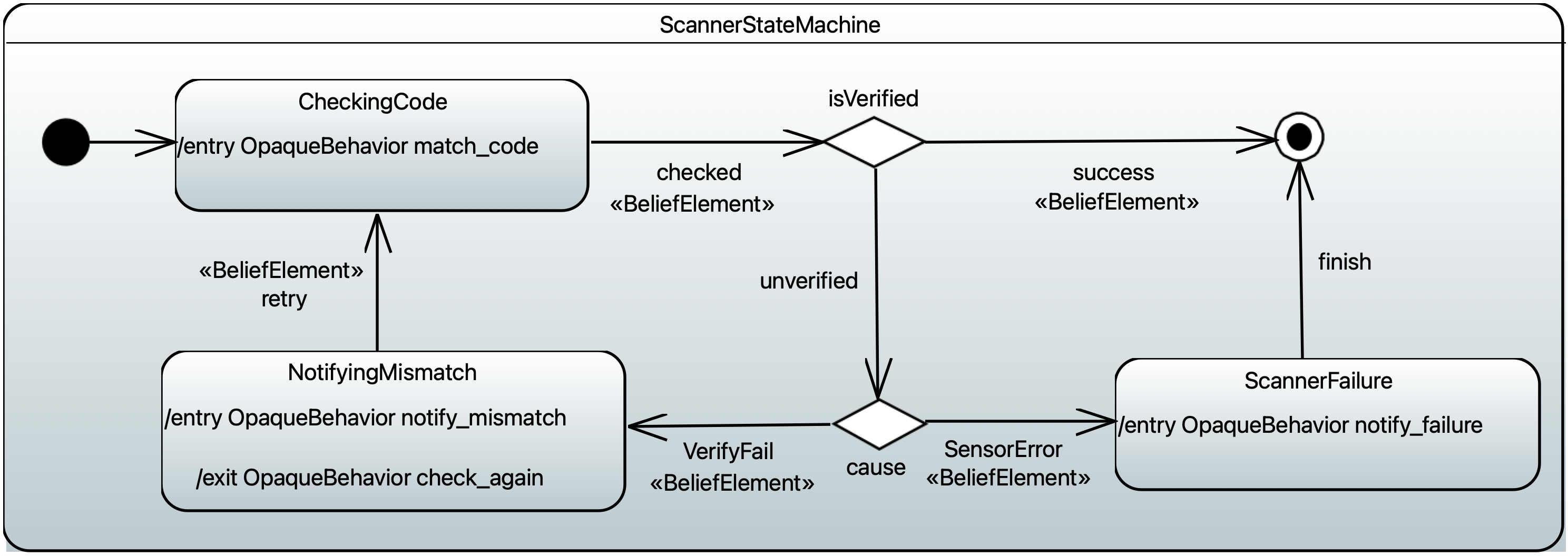}}
\caption{A UML state machine for a device's scanner depicting medicine barcode scanning procedure.}
\label{fig:scanner-sm}
\end{figure*}

\subsubsection{Device Battery Behavior}
The state machine of a dispenser's battery is depicted in \Cref{fig:battery-sm}. 
Depending on the availability of the power source, the battery state machine has states \emph{ChargingBattery} and \emph{UsingBattery}. 
Since most modern dispensers come with a rechargeable battery, \emph{ChargingBattery} occurs when a battery is supplied with a power source for charging. 
If the power supply is connected, the battery is in \emph{ChargingBattery} state. 
During this state, if the battery is not charged and a power failure occurs, it generates \emph{NoPower} event. 
When the battery is fully charged and the power supply is unplugged, the dispenser utilizes the battery. 
This state is represented as \emph{UsingBattery}. 
When the battery level becomes low, the device generates \emph{LowBattery} event indicating the need for a power supply. 
If the battery runs out of charging, the device shuts down after raising the \emph{DeadBattery} event. 
All transitions with \guillemetleft BeliefElement\guillemetright~stereotype indicate uncertainties.

\subsubsection{Device Scanner Behavior}
The state machine of a barcode scanner is shown in \Cref{fig:scanner-sm}. 
The scanner behavior involves \emph{CheckingCode}, \emph{ScannerFailure}, and \emph{NotifyingMismatch} states. 
Its operation starts with the state \emph{CheckingCode} in which the barcode on the medicine roll is matched to check if the medicine is intended for the specific patient. 
After completing barcode matching, it ends up in either successful or unverified situations. 
If medicines are scanned and verified successfully, the process is completed. 
In the case of unverified medicines, there can be one of the two situations, i.e., scanning failed due to mismatch or scanner error. 
This leads to two possible states: \emph{NotifyingMismatch}, corresponding to a \emph{VerifyFail} event, and \emph{ScannerFailure}, corresponding to a \emph{SensorError} event.
In the event of a mismatch, the device alerts about the discrepancy and prompts the patient to change the medication roll.
Each state transition involving uncertainties is modeled with the stereotype \guillemetleft BeliefElement\guillemetright.

\subsubsection{Device Camera Behavior}
\Cref{fig:camera-sm} illustrates the state machine for the dispenser's camera, containing the states \emph{RecognizingFace}, \emph{NotifyingMismatch}, and \emph{CameraFailure}.
The camera begins its operation with \emph{RecognizingFace} state, where it performs facial recognition of the patient to confirm that the medicine intake is by the intended patient. 
Upon completing the facial recognition process, the verification succeeds or remains unverified. 
In case of successful verification, the device starts the dispensing procedure. 
If facial recognition remains unverified, this could be due to either a mismatch or a camera failure.
These scenarios result in two states: \emph{NotifyingMismatch}, which corresponds to a \emph{VerifyFail} event, and \emph{CameraFailure}, which corresponds to a \emph{SensorError} event.
In the case of a mismatch, it encourages the patient to attempt again, while simultaneously notifying the IoT-based healthcare application about the discrepancy.
As transitions among camera states may have uncertainties, the stereotype \guillemetleft BeliefElement\guillemetright~is applied to these transitions.

\subsection{Model Interpretation}
Once device environment models are developed (using methodology discussed in \Cref{sec:envmodels}), our approach processes these models to make them executable for the simulation.
First, we take the UML class diagram of the environment and create its internal representation. 
This involves interpreting environment classes, properties, signal receptions, applied stereotypes, and owned behaviors. 
Using the environment profile stereotypes applied to the classes, we distinguish between environment components and the DT part. 
During this phase, we also identify uncertain elements modeled with \textcolor{black}{stereotypes (such as \guillemetleft BeliefElement\guillemetright) from} UUP. 
From each uncertain element, we extract its associated probability, which is required during environment simulation. 

In subsequent steps, we create instances of all environment classes representing an environment component. 
To instantiate the class diagram, we utilize OCL constraints to determine the number of instances to create and correct property value ranges. 
We create class instances considering the allowed number specified in  OCL constraints, for example \emph{C4} in \Cref{lst:env-constraints}. 
For the class properties, we assign their values randomly from the ranges specified in the OCL constraints.
To illustrate, for constraint \emph{C1} in \Cref{lst:env-constraints}, we randomly generate a battery \emph{level} value between 0 and 100. 
The instance model generated during this process represents a specific environment configuration. 

The next step is to interpret the owned behavior (i.e., state machine) of each environment class. 
This requires traversing state machines to identify states, transitions, triggers, events, applied stereotypes, and submachines. 
For each state featuring \emph{OpaqueBehavior}, namely \emph{Entry}, \emph{Exit}, and \emph{DoAcivity}, we extract the languages used to specify a specific behavior and the corresponding bodies. 
If a state features a submachine as an internal behavior, we load it following the same procedures and associate it with the state. 
\textcolor{black}{As we traverse the transitions, we extract each transition's associated information including triggers and events. 
In cases where a transition is annotated with uncertain stereotypes (like \guillemetleft BeliefElement\guillemetright), we extract its corresponding probability value specified as a part of measurement kind (as discussed in~\Cref{sec:devbehmodel}). 
Moreover, we retrieve the selected probability distribution and its property values. 
}
After traversing state machines of all environment classes, we establish an association with the instance model. 
Specifically, for each environment class possessing owned behavior, we link the corresponding behavior with the instance of that class.
Upon concluding the interpretation of the environment models, these models are ready to be utilized for the simulations. 

\begin{algorithm}[!ht]
\caption{\textcolor{black}{Environment simulation}}\label{algo:envsim}
  \DontPrintSemicolon
  \SetAlgoLined
  \SetKwComment{Comment}{$\triangleright$ }{}
  \SetKwFunction{FExecuteBehavior}{ExecuteBehavior}
  \SetKwFunction{FExecuteState}{ExecuteState}
  \SetKwFunction{FExecuteTransition}{ExecuteTransition}
  \SetKwFunction{FFindTransition}{FindTransition}
  \textcolor{black}{
  \KwIn{$PD$\Comment*[r]{ProbabilityDistribution}}  
  \SetKwProg{Fn}{Function}{:}{\KwRet}
  \Fn{\FExecuteBehavior{$M$}}{
    $s_i \gets getInitialState(S)$\;
    $t_x \gets$ \FFindTransition{$s_i$}\;
    \FExecuteTransition{$t_x$}\;
  }
  \SetKwProg{Pn}{Function}{:}{\KwRet $t_x$}
  \Pn{\FFindTransition{$s_x$}}{
    $outgoings \gets getOutgoings(s_x)$\;
    $LH_{t} \gets \emptyset$\Comment*[r]{Likelihood list}
    \ForEach{$ t_o \in outgoings $}
    {
    $e_x \gets getEvent(t_o)$\;
    $p_o \gets checkUncertainty(t_o)$\;
    $p_n \gets getLikelihood(e_x, p_o, PD)$\;
    $LH_{t} \gets t_o, p_n$\;
    }
    $t_x \gets determineTransition(LH_{t})$\;
  }
  \SetKwProg{Pn}{Function}{:}{\KwRet}
  \Pn{\FExecuteTransition{$t_x$}}{
    $s_{x+1} \gets getTarget(t_x)$\;
    $e_x \gets getEvent(t_x)$\;
    $notifyDT(e_x)$\;
    \FExecuteState{$s_{x+1}$}\Comment*[r]{Next state}
  }
  \SetKwProg{Pn}{Function}{:}{\KwRet}
  \Pn{\FExecuteState{$s_x$}}{
        \eIf{$isFinal(s_x)$}
        {\KwRet \Comment*[r]{Finish}}
        {$runOpaqueBehaviors(s_x)$\;
        \eIf{$hasSubmachine(s_x)$}
        {
        $sm \gets getSubmachine(s_x)$\;
        \FExecuteBehavior{$sm$}\Comment*[r]{$\parallel$}
        }
        {
        $t_x\gets$ \FFindTransition{$s_x$}\;
        \FExecuteTransition{$t_x$}\;
        }
        }
  }
  }
\end{algorithm}

\subsection{Environment Simulation}\label{sec:envsimulation}
Our approach takes environment models interpreted in previous steps to create environment simulations.
\textcolor{black}{
Since the simulation involves executing state machines, we first define a formalism for state machines with uncertainties. 
This formalism is then utilized to demonstrate the simulation process. 
We denote a state machine as a tuple \(M = (S, s_i, s_f, E, T)\), where: 
\begin{itemize}
\item $S = \{s_i, s_1, s_2, \ldots, s_f\}$ is a finite set of states. 
\item $s_i$ and $s_f \in S$ are initial and final states, respectively. 
\item $E = \{e_1, e_2, \ldots, e_n\}$ is a finite set of events that cause transitions.
\item $T: s_x \xrightarrow[]{\text{$e_k/p$}} s_y$ is a transition function, where $s_x, s_y \in S$. For each transition $T(s_x, e_k, s_y)$, the system will transition from state $s_x$ to state $s_y$ when event $e_k$ occurs with probability $p$. A transition with $p=1$ indicates a deterministic transition while a transition with $p=[0,1)$ indicates an uncertain transition. 
\end{itemize}
}

At the beginning of this process, we start with executing the device state machine \textcolor{black}{(\Cref{algo:envsim}, lines: 2--6)}. 
Starting from an initial pseudostate \textcolor{black}{($s_i$), we first examine all outgoing transitions and determine which transition to take (\Cref{algo:envsim}, lines: 7--17). 
For each outgoing transition, we check the applied uncertain stereotype and identify its associated event $e_k$ and probability $p$. 
If an outgoing transition does not have an applied uncertain stereotype, it is treated as a deterministic transition with a high likelihood, i.e., $p=1$. 
Subsequently, we use the specified probability distribution (\Cref{algo:envsim}, line: 1) to determine the likelihood of an event's occurrence. 
Upon determining the likelihood for all outgoing transitions, we select the transition with the highest likelihood. 
We then proceed with executing the transition associated with the event having the highest likelihood of occurrence (\Cref{algo:envsim}, lines: 18--23). 
While transitioning to the next state, we identify the target state and execute the associated signal event. 
Executing signal events involves generating events and dispatching them to the DT. 
Following this, we execute the next target state (\Cref{algo:envsim}, lines: 24--37). 
On entering a state,  we sequentially execute all Opaque Behaviors. 
If an \emph{Entry} behavior is available, we begin with its execution. 
Specifically, we run code contained within the body, which, in our context, must be written in Python. 
After executing \emph{Entry} behavior, if \emph{DoActivity} behavior is modeled for the state, we run code inside the body as a Python executable script. 
Once the \emph{DoActivity} behavior execution finishes, we execute the \emph{Exit} behavior similarly. 
Next, if a state possesses an internal behavior in the form of a submachine, we initiate the execution of the sub-state machine concurrently (\Cref{algo:envsim}, lines: 29--31). 
When the execution of a state is completed, we determine the next transition to take and execute it (\Cref{algo:envsim}, lines: 32--35). 
This process of executing state behaviors and transitioning to another state continues during the simulation. 
The execution of a behavior finishes when the final state is reached (\Cref{algo:envsim}, lines: 25--26). 
}

\textcolor{black}{
For the dispenser example shown in \Cref{fig:dev-dt-sm}, the transition from $s_i$ to \emph{SettingUp} state does not have an uncertain stereotype. 
Therefore, we take the outgoing transition to the \emph{SettingUp} state ($s_1$). 
Now, from the state $s_1$, there are two outgoing transitions and both have \guillemetleft BeliefElement\guillemetright~stereotype. 
One of the transitions ($T_1$) named \emph{initialized} leads to \emph{CheckingRoll} state ($s_2$) and have associated event $e_1: CheckRoll$. 
The other transition ($T_2$) named \emph{shutdown} leads to \emph{ShuttingDown} state (say $s_6$) and have associated event $e_2: ShutDown$. 
If for example the probabilities of events $e_1$ and $e_2$ are 0.8 and 0.2 respectively, both transitions can be denoted as follows. 
$T_1: s_1 \xrightarrow[]{\text{$e_1/0.8$}} s_2$. 
$T_2: s_1 \xrightarrow[]{\text{$e_2/0.2$}} s_6$. 
Now which transition to take depends on the value sampled from a probability distribution. 
For this purpose, we draw the likelihood of both events from a probability distribution and compare it with given probabilities. 
If the comparison favors event $e_1$, we select transition $T_1$, otherwise, we choose transition $T_2$. 
For instance, if transition $T_1$ is to be executed, its corresponding event will be dispatched to the DT and the state will be transitioned to $s_2$ (\emph{CheckingRoll}). 
Similarly, $s_2$ and the subsequent states will continue to execute until the device reaches its final state ($s_6$) upon shutdown. 
}

During the execution of each state's owned behaviors (i.e., \emph{Entry}, \emph{DoActivity}, \emph{Exit}, or \emph{Submachine}), if there are any changes to the property values of any environment components, we update the instance model with the current values. 
For instance, in \emph{load\_configs} \emph{DoActivity}, the default device configurations are loaded. 
In response to this, we also update the instance model. 
By doing so, the instance model maintains the updated configurations of the environment.
Furthermore, for states involving changes related to user interactions, e.g., modifications of device alarm settings, we randomly generate values based on ranges defined in OCL constraints and accordingly update the instance model. 
During the device behavior simulation, the environment components' behavior also executes concurrently, and the instance model continuously adapts to the updated environment configurations.

\subsection{Operating DT with Simulated Environment during Testing}
To generate and operate a device's DT, we follow the previous model-based approach~\cite{sartaj2024modelbased}. 
Specifically, we use device input configurations to generate an instance model and a state machine to create an executable model representing a DT. 
To operate DT and its integration with an IoT-based healthcare application, we create a DT server and APIs (Application Programming Interfaces).

During testing, a DT integrated with an IoT-based healthcare application begins its operation.
As a DT operates, it transitions through various states while executing behaviors, such as the \emph{Dispensing} state of medicine dispensers. 
The DT behavior states correspond to the device behavior in environment models. 
Therefore, during environment simulation, the DT receives various signal events asynchronously. 
The DT responds to these events by adjusting its state or properties in the instance model as required. 

\subsection{\textcolor{black}{\approach{} Implementation}}
We implemented \approach{} using Java and Python. 
To load environment models developed in UML, we utilized Java, leveraging the Eclipse Modeling Framework (EMF)~\cite{steinberg2009emf} Java library. 
This choice was made as the Python equivalent, specifically PyUML2~\cite{pyuml}, is currently in its initial development stages and lacks comprehensive support for UML models. 
As the DT generation approach~\cite{sartaj2024modelbased} was implemented in Python (available at~\cite{repo}), we developed the remaining components of our approach in Python as well to ensure seamless compatibility. 
To communicate with Java code from Python, we used Py4J~\cite{py4j}, which allows dynamic access and manipulation of Java objects. 
For the creation of internal model representations (e.g., model instantiations), we relied on PyEcore~\cite{pyecore}---a Python variant of EMF.

\section{Evaluation}\label{sec:evaluation}
\textcolor{black}{
In this section, we present the evaluation of the \approach{}, conducted using three real-world medical devices connected to an industrial IoT-based healthcare application. 
We initially define the overall goal of the evaluation and the research questions to achieve the goal. 
Following that, we present case studies used in our evaluation, the experiment setup and execution, the analysis of experiment data, results corresponding to each research question, and the discussion of results. 
}

\subsection{\textcolor{black}{Goal and Research Questions}}
\textcolor{black}{
Our evaluation goal is to demonstrate the \approach{}'s applicability for industry practitioners. 
To apply or customize \approach{} for a particular device, it is necessary to determine the probability distribution that best represents the environmental uncertainties. 
Nevertheless, practitioners typically lack knowledge regarding the precise probability distribution followed by a particular device. 
This is because various vendors supply diverse types of medical devices, and the exact probability distribution each device follows is typically either undisclosed or unknown. 
Therefore, we analyze the applicability of \approach{} in determining the relevant probability distributions. 
Furthermore, as \approach{} is designed to support the testing of IoT-based healthcare applications, an important consideration for industry practitioners is to examine whether the generated simulations adequately represent uncertain environmental scenarios. 
During the environment simulation, \approach{} executes behavioral models (i.e., state machines) based on various probability distributions (\Cref{sec:envsimulation}). 
Due to these probabilistic model executions, several states and transitions may execute repetitively, leaving some unexplored. 
This might lead to the repetitive generation of similar uncertain scenarios. 
From a testing perspective, this could be undesirable as similar uncertainties may likely reveal the same types of faults. 
Given that industry practitioners are typically interested in analyzing coverage of model elements and the diversity of scenarios for testing~\cite{hemmati2013achieving}. 
Therefore, considering the role of environment models and probability distributions in simulations, we identify two key parameters for evaluating the applicability of \approach{}: (i) the coverage of environment models, and (ii) the diversity of uncertain scenarios. 
Coverage analysis allows us to examine the model elements covered during probabilistic simulation. 
Meanwhile, diversity analysis enables us to investigate the variety of uncertain scenarios generated during the simulation, correlating with different probability distributions. 
}

\textcolor{black}{
Regarding \approach{}'s comparison with baseline approaches, we identified a few works on environment simulations~\cite{llopis2023modeling,iqbal2015environment,haris2019sensyml}. 
However, these approaches are designed for a particular system or domain, e.g., smart offices~\cite{llopis2023modeling}. 
Adapting these approaches to medical devices would require extensive customization, which includes tailoring modeling methodology and a specialized approach for handling uncertainties. 
Given the differences in domain and environment, tailoring and implementing these approaches would also introduce considerable implementation bias. 
This bias could emerge from differences in programming languages (like Python or Java), variations in technologies or libraries, and differences in platforms (such as MacOS or Linux). 
Such biases could pose a potential threat to the internal validity of the evaluation. 
Hence, employing a comparison baseline for this evaluation is not feasible. 
}

\textcolor{black}{
Based on the evaluation goal, we formulate the following research questions (RQs). 
}

\begin{itemize}
    \item[] \colorbox{teal!10!white}{\textbf{RQ1:}} How much coverage of environment models is achieved with various probability distributions?
    \\\textcolor{black}{Given that various probability distributions can influence model coverage during simulation, this RQ aims to analyze the extent to which the selection of a probability distribution impacts model coverage. This aspect is important in determining probability distributions that can sufficiently cover environment models during simulation.} 
    \item[] \colorbox{teal!10!white}{\textbf{RQ2:}} What is the diversity of uncertain scenarios generated for the DT during the probabilistic environment simulations?
    \\\textcolor{black}{Considering the significance of diverse scenarios in a testing context, this RQ aims to analyze the impact of various probability distributions on the diversity of uncertain scenarios generated during simulation. This aspect is essential in determining probability distributions that sufficiently explore diverse environmental uncertain scenarios necessary for testing.} 
\end{itemize}

\subsection{\textcolor{black}{Case Studies for Experiment}}
We used three types of medicine dispensers---namely, \emph{Karie}, \emph{Medido}, and \emph{Pilly}---which the health department of Oslo City provided as a part of the experimental setup. 
These medicine dispensers have been integrated with a real-world IoT-based healthcare application and are currently used across multiple counties in Norway. 
Following, we provide a detailed overview of the key features of each dispenser.

\subsubsection{Karie}
Karie~\cite{karie} is an automatic medicine dispenser that provides users with various customizable settings such as alarm melody, volume, language, and brightness. 
These settings can be adjusted either through a user-friendly graphical interface designed for patients or via an IoT-based healthcare application (usually operated by caregivers). 
Karie automatically retrieves the medication plan from the IoT-based healthcare application, follows the plan, rings an alarm to remind the patient, and dispenses medicine. 
Moreover, it conveys essential information to the IoT application, such as notifications about missed and taken doses. 
This feature ensures that caregivers remain informed about the patient's medication adherence, aiding in effective health monitoring.

\subsubsection{Medido}
Medido~\cite{medido} is an automatic medication dispenser that offers an interface for IoT-based healthcare applications, enabling caregivers or medical professionals to modify settings. 
It supports alarm and medication configurations and is multilingual with support for three languages. 
At its core, Medido fetches medication schedules from the IoT-based healthcare application, dispenses doses per this schedule, and communicates any issues with medicine delivery, intake doses, and forgotten medication.

\subsubsection{Pilly}
Pilly~\cite{pilly} is a lightweight, low-end medication dispenser that is ideally suited for travel purposes. 
It offers basic medication features like setting the alarm duration and volume. 
All settings can only be modified via an IoT-based healthcare application. 
It supports a fixed medication plan with a predetermined number of doses. 
It loads a medication plan from an IoT-based healthcare application, dispenses medicine, and informs about medication intake and missed doses. 

\begin{table}[h]
\caption{Statistics of environment models for each case study.}\label{tab-modstats}
\begin{tabular}{@{}llll@{}}
    \toprule
    \textbf{Model Elements} & \textbf{Karie}  & \textbf{Medido} & \textbf{Pilly}\\
    \midrule
    Classes    &  8   &  7  &  5  \\
    Properties    &  11   &  10  &  10  \\
    Class Stereotypes    &  9   &  9  &  7  \\ 
    Signal Receptions    &  13   &  12  &  8  \\
    OCL Constraints  &  12   &  11  &  9  \\ 
    \arrayrulecolor{black!20}\midrule[0.1pt]
    State Machines (SM)   &  8   &  7  &  4  \\
    SM Stereotypes    &  6   &  6  &  5  \\ 
    States    &  32   &  23  &  12  \\
    Transitions    &  70   &  51  &  26  \\
    All Events    &  54   &  39  &  19  \\
    Uncertain Events    &  40   &  31  &  13  \\
    Opaque Behaviors    &  46   &  36  &  22  \\
    \botrule
\end{tabular}
\end{table}

\subsection{\textcolor{black}{Experiment Setup}}
We developed environment models for Karie, Medido, and Pilly devices using the modeling methodology presented in \Cref{sec:approach}. 
Mainly, we used the environment modeling profile, signal events model library, and UUP. 
\Cref{tab-modstats} presents modeling statistics for each case study, i.e., Karie, Medido, and Pilly. 
The class diagram for Karie encompasses eight classes, 11 properties of primitive types, nine unique stereotypes, 13 signal receptions, and 12 OCL constraints. 
Similarly, the Medido class diagram includes seven classes, 10 primitive properties, nine unique stereotypes, 12 signal receptions, and 11 OCL constraints. 
Lastly, the Pilly class diagram consists of five classes, 10 properties of primitive types, seven unique stereotypes, eight signal receptions, and nine OCL constraints.

For the Karie device's behavioral models, we designed eight state machines for each environmental component possessing behavior. 
These state machines include six unique stereotypes, 32 states, 70 transitions, 54 total number of events, 40 uncertainty-associated events, and 46 Opaque Behaviors (i.e., \emph{Entry}, \emph{DoActivity}, and \emph{Exit}).
In the case of the Medido device, we developed seven state machines each reflecting an individual environmental component's behavior. 
The Medido device state machines encompass six unique stereotypes, 23 states, 51 transitions, a total of 39 events, 31 events related to uncertainty, and 36 Opaque Behaviors, namely \emph{Entry}, \emph{DoActivity}, and \emph{Exit}.
Lastly, for the Pilly device's behavioral models, we created four state machines corresponding to each environment component with behavior. 
The state machines for the Pilly device comprise five unique stereotypes, 12 states, 26 transitions, 19 total number of events, 13 uncertain events, and 22 Opaque Behaviors, specifically, \emph{Entry}, \emph{DoActivity}, and \emph{Exit}). 
In all state machines, we determined the probabilities associated with uncertain elements based on our hands-on experience with these devices and insights gained from professionals in the Oslo City healthcare department.

To specify probability distributions for RQ1 and RQ2, we selected all distributions from the UUP \emph{Probability} model library~\cite{zhang2019uncertainty}. 
This selection includes \emph{NormalDistribution}, \emph{BinomialDistribution}, \emph{BernoulliDistribution}, \emph{ExponentialDistribution}, \emph{GammaDistribution}, \emph{PoissonDistribution}, \emph{UniformDistribution}, \emph{GeometricDistribution}, \emph{TriangularDistribution}, and \emph{LogarithmicDistribution}.  
In our experiments, we individually configured each distribution, as our objective was to analyze the impact of various probability distributions on the environment simulations. Note that real distributions of various uncertainties are unknown in our context due to the lack of data (i.e., due to confidentiality concerns). Moreover, domain experts cannot estimate such distributions. Nonetheless, in this paper, we aim to demonstrate that with our approach, one can model various uncertainties and simulate them in the environment of a DT.

\subsection{\textcolor{black}{Experiment Execution}}
We used the environment models created for each case study (\textit{Karie}, \textit{Medido}, and \textit{Pilly}) to execute simulations. 
The environment simulations in our approach rely on probabilities, introducing a randomness factor.
To account for this, we repeated the experiment 30 times, adhering to the standard practice~\cite{arcuri2011practical} for experiments with stochastic elements. 
Furthermore, the state machines of the devices used in our experiments might contain loops (for instance, \emph{ConnectivityStateMachine} shown in \Cref{fig:conn-sm}). 
These loops in state machines can lead to repeated executions of the same model elements while bypassing probabilistic effects, potentially resulting in 100\% coverage in every experimental run. 
To avoid such scenarios, we configured simulations to exercise each model element only once and according to the likelihood determined from probability distributions. 
For executing experiments, we utilized a machine equipped with a macOS operating system, an 8-core CPU, and 24 GB RAM.

\subsection{Data Analysis}
\textcolor{black}{
\Cref{tab-evaldesign} presents analysis context, parameters, and metrics corresponding to each RQ. 
For RQ1, our analysis focuses on applicability, considering parameters such as probability distribution (PD) and coverage. 
}
To evaluate model coverage for RQ1, we monitored the execution traces of the behavioral elements within the environment models.
Specifically, in each simulation, we examined various model elements, including the number of class instances created or updated, the frequency of states executed, the number of \emph{Entry}/\emph{Exit}/\emph{DoActivity} behaviors executed, the count of transitions taken, and the number of events generated. 
The overall coverage was calculated based on the proportion of model elements traced during various executions to the total number of model elements. 
A high coverage percentage signifies that a substantial amount of model elements are exercised/covered during environment simulations.

\textcolor{black}{
For RQ2, we analyze applicability by considering parameters like PD and diversity. 
}
To analyze the diversity of uncertain events for RQ2, we used Simpson’s diversity measure~\cite{simpson1949measurement}.  
Given that the simulation of uncertain environmental events is based on the probabilities of their occurrences, we adhered to the guidelines provided by Roswell et al.~\cite{roswell2021conceptual} for selecting this measure.
A diversity value close to 1 demonstrates high diversity, and a value close to 0 corresponds to low diversity. 
Furthermore, we calculated the mean and standard deviation (STD) across all repetitions to present aggregated results for RQ1 and RQ2.

\begin{table}[!t]
\caption{\textcolor{black}{RQs' mapping with analysis context, parameters, and metrics. }}\label{tab-evaldesign}
\textcolor{black}{
\begin{tabular}{@{}lllp{2.4cm}@{}}
    \toprule
    \textbf{RQ} & \textbf{Analysis}  & \textbf{Parameters} & \textbf{Metrics}\\
    \midrule
    RQ1    &  Applicability   &  PD, coverage  &  \% coverage, Mean, STD  \\
    RQ2    &  Applicability   &  PD, diversity  &  Simpson diversity, Mean, STD  \\
    \botrule
\end{tabular}
}
\end{table}

\begin{figure*}[htbp]
\centerline{\includegraphics[width=\textwidth, keepaspectratio]{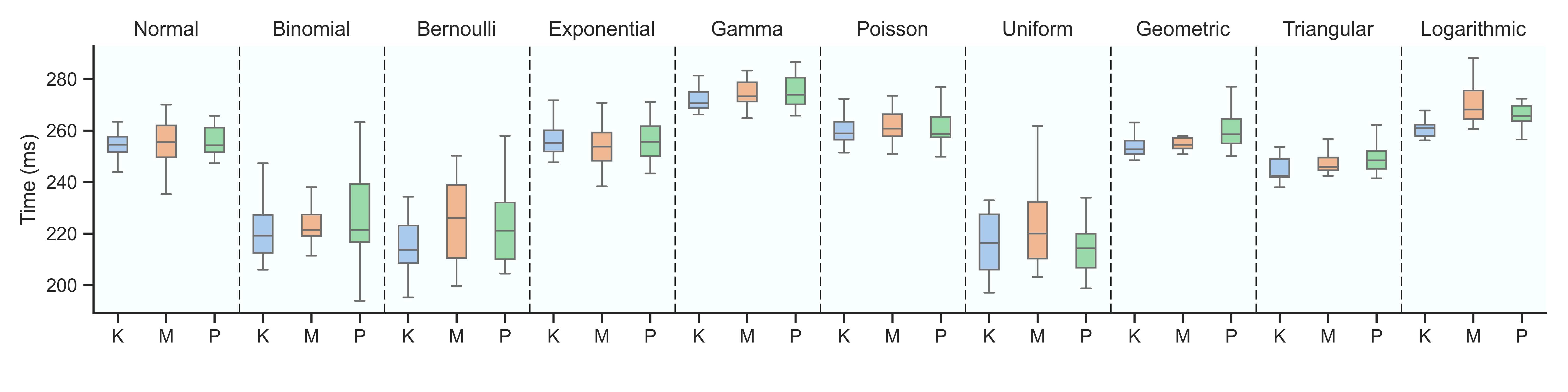}}
\caption{Simulation time measured across all repetitions for each case study, where K, M, and P denote Karie, Medido, and Pilly, respectively.}
\label{fig:sim-time}
\end{figure*}

\begin{table*}[!t]
	\centering
	\noindent
	\caption{RQ1 results showing model coverage achieved with different probability distributions. These results are presented in terms of mean percentage coverage and standard deviation (STD) computed for all experiment runs. Values with high mean coverage are marked in green, while low STD values are highlighted in red.}
	\begin{tabular}{@{}l l l l l l l@{}}\toprule
		\multicolumn{1}{l }{\textbf{}} & \multicolumn{2}{c }{\textbf{Karie}} & \multicolumn{2}{c }{\textbf{Medido}}& \multicolumn{2}{c }{\textbf{Pilly}} \\ 
		\cmidrule(lr){2-3}
		\cmidrule(ll){4-5} \cmidrule(ll){6-7}
		\multicolumn{1}{ l }{\textbf{Distribution}} & \textbf{Mean} & \textbf{STD}& \textbf{Mean} & \textbf{STD}& \textbf{Mean} & \textbf{STD}\\ 
		\cmidrule(lr){1-1}
		\cmidrule(lr){2-3}
		\cmidrule(ll){4-5}\cmidrule(ll){6-7}
		\multicolumn{1}{ l }{\textbf{Normal}} & \cellcolor{green!5}60.8\% & 1.10& \cellcolor{green!5}61.01\% & 0.97& \cellcolor{green!5}60.84\% & 1.08\\
        \multicolumn{1}{ l }{\textbf{Binomial}} & 43.25\% & 5.01& 42.48\% & 5.20& 43.1\% & 5.88\\
        \multicolumn{1}{ l }{\textbf{Bernoulli}} & 43.15\% & 4.61& 44.30\% & 5.90& 44.74\% & 5.67\\
        \multicolumn{1}{ l }{\textbf{Exponential}} & \cellcolor{green!5}61.27\% & \cellcolor{green!5}0.36& \cellcolor{green!5}61.37\% & \cellcolor{green!5}0.34& \cellcolor{green!5}61.42\% & \cellcolor{green!5}0.32\\
        \multicolumn{1}{ l }{\textbf{Gamma}} & 45.92\% & 2.64& 47.25\% & 2.70& 46.58\% & 2.95\\
        \multicolumn{1}{ l }{\textbf{Poisson}} & \cellcolor{green!5}60.3\% & 1.04& \cellcolor{green!5}60.16\% & 1.31&\cellcolor{green!5}60.49\% & 0.98\\
        \multicolumn{1}{ l }{\textbf{Uniform}} & \cellcolor{red!5}34.88\% & 3.31& \cellcolor{red!5}36.16\% & 3.03& \cellcolor{red!5}35.99\% & 3.24\\
        \multicolumn{1}{ l }{\textbf{Geometric}} & \cellcolor{green!5}61.57\% & \cellcolor{green!5}0.15& \cellcolor{green!5}61.55\% & \cellcolor{green!5}0.33& \cellcolor{green!5}61.55\% & \cellcolor{green!5}0.29\\
        \multicolumn{1}{ l }{\textbf{Triangular}} & \cellcolor{green!5}60.81\% & 0.76& 59.76\% & 1.51& \cellcolor{green!5}60.26\% & 1.13\\
        \multicolumn{1}{ l }{\textbf{Logarithmic}} & \cellcolor{green!5}60.78\% & 1.79& \cellcolor{green!5}61.14\% & 0.96& \cellcolor{green!5}61.25\% & 0.54\\
		\bottomrule
	\end{tabular}
	\label{tab-rq1results}
\end{table*}

\begin{table*}[!t]
	\centering
	\noindent
	\caption{RQ2 results for the diversity of uncertain scenarios w.r.t. each probability distribution. These results are presented in terms of mean percentage diversity and standard deviation (STD) calculated across all experiment runs. High diversity mean values are indicated in green, while low STD values are marked in red. }
	\begin{tabular}{@{}l l l l l l l@{}}\toprule
		\multicolumn{1}{l }{\textbf{}} & \multicolumn{2}{c }{\textbf{Karie}} & \multicolumn{2}{c }{\textbf{Medido}}& \multicolumn{2}{c }{\textbf{Pilly}} \\ 
		\cmidrule(lr){2-3}
		\cmidrule(ll){4-5} \cmidrule(ll){6-7}
		\multicolumn{1}{ l }{\textbf{Distribution}} & \textbf{Mean} & \textbf{STD}& \textbf{Mean} & \textbf{STD}& \textbf{Mean} & \textbf{STD}\\ 
		\cmidrule(lr){1-1}
		\cmidrule(lr){2-3}
		\cmidrule(ll){4-5}\cmidrule(ll){6-7}
		\multicolumn{1}{ l }{\textbf{Normal}} & \cellcolor{green!5}0.60 & 0.05& \cellcolor{green!5}0.62 & \cellcolor{green!5}0.02& \cellcolor{green!5}0.61 & 0.03\\
        \multicolumn{1}{ l }{\textbf{Binomial}} & 0.38 & 0.27& 0.30 & 0.32& 0.37 & 0.29\\
        \multicolumn{1}{ l }{\textbf{Bernoulli}} & 0.32 & 0.30& 0.36 & 0.26& 0.43 & 0.24\\
        \multicolumn{1}{ l }{\textbf{Exponential}} & \cellcolor{green!5}0.62 & \cellcolor{green!5}0.01& \cellcolor{green!5}0.61 & 0.06& \cellcolor{green!5}0.62 & \cellcolor{green!5}0.01\\
        \multicolumn{1}{ l }{\textbf{Gamma}} & 0.40 & 0.23& 0.30 & 0.24& 0.32 & 0.26\\
        \multicolumn{1}{ l }{\textbf{Poisson}} & \cellcolor{green!5}0.61 & 0.03& \cellcolor{green!5}0.62 & 0.03& \cellcolor{green!5}0.61 & 0.03\\
        \multicolumn{1}{ l }{\textbf{Uniform}} & \cellcolor{red!5}0.02 & 0.10& \cellcolor{red!5}0.08 & 0.19& \cellcolor{red!5}0.05 & 0.13\\
        \multicolumn{1}{ l }{\textbf{Geometric}} & \cellcolor{green!5}0.62 & \cellcolor{green!5}0.02& 0.56 & 0.12& 0.53 & 0.16\\
        \multicolumn{1}{ l }{\textbf{Triangular}} & \cellcolor{green!5}0.61 & 0.04& \cellcolor{green!5}0.60 & 0.07& \cellcolor{green!5}0.62 & \cellcolor{green!5}0.02\\
        \multicolumn{1}{ l }{\textbf{Logarithmic}} & \cellcolor{green!5}0.62 & \cellcolor{green!5}0.02& \cellcolor{green!5}0.61 & \cellcolor{green!5}0.02& \cellcolor{green!5}0.62 & 0.03\\
		\bottomrule
	\end{tabular}
	\label{tab-rq2results}
\end{table*}

\subsection{Experiment Results}
Throughout the execution of the experiment, we recorded the simulation time for each repetition. 
\Cref{fig:sim-time} presents boxplots illustrating each device's simulation time (expressed in milliseconds) for various probability distributions across all experimental runs. 
It is important to note that the recorded simulation time excludes the waiting periods required in various states of the devices' state machines, as our focus was on analyzing the core runtimes of different model executions.
Taking an overall view of the boxplots, it is evident that the simulation time mainly falls within the range of 200-300 milliseconds. 
For certain probability distributions, such as Normal and Exponential, simulation times mostly exceed 240 milliseconds, whereas for others, like Binomial and Bernoulli, they remain below 240 milliseconds. 
Upon analyzing the coverage results (as detailed in \Cref{sec:rq1res}) for RQ1, we observed that simulations of longer duration also resulted in increased coverage, suggesting more model elements were exercised during the simulations. 

The following sections discuss experiment results, addressing each RQ individually.

\subsubsection{RQ1 Results}\label{sec:rq1res}

To answer RQ1 targeting environment model coverage, \Cref{tab-rq1results} presents coverage results for the Karie, Medido, and Pilly devices computed across all experiment repetitions and for each probability distribution.

In the case of the Karie device, an average model coverage exceeding 60\% using Normal, Exponential, Poisson, Geometric, Triangular, and Logarithmic distributions. 
Among these results, the model coverages obtained through the Exponential and Geometric distributions exhibit low STD values. 
This suggests a lower variability within coverage results, indicating a more consistent performance across all runs.
Moreover, the results associated with the remaining distributions indicate low model coverage with high STD. 
Notably, the coverage results for the Uniform distribution show the lowest coverage, approximately 34\%.

Similarly, for the Medido device, an average model coverage achieved was approximately 61\%, utilizing Normal, Exponential, Poisson, Geometric, and Logarithmic distributions. 
After analyzing STD values for these results, it is evident that the model coverages obtained through the Exponential and Geometric distributions demonstrate lower STD values. 
This indicates more consistent coverage results throughout all experimental runs.
In addition, the other distributions demonstrated low model coverage and high STD, particularly the Uniform distribution which yielded the lowest coverage at approximately 36\%.

Lastly, in the case of the Pilly device, both the set of Normal, Poisson, and Triangular distributions and the set of Exponential, Geometric, and Logarithmic distributions yielded an average model coverage of approximately 60\% and 61\%, respectively. 
Among these distributions, the Exponential and Geometric with high model coverage show low STD values. 
This signifies a more consistent coverage outcome across all experimental runs.
Like the Karie and Medido devices, the Uniform distribution also exhibited the lowest coverage at approximately 36\% for the Pilly device.

An analysis of the overall results indicates that the maximum coverage achieved is approximately 61\%. 
Upon observing simulation traces, we perceived that this is primarily due to the uncertain nature of the environment simulations, which seem to resemble realistic conditions. 
For example, Karie's medicine scanner (as shown in \Cref{fig:scanner-sm}) rarely fails, while mismatches may occur quite often. 
Achieving a full 100\% coverage would imply deterministic environment simulations, where every model element is always covered. 
This would mean that each scenario executes with absolute certainty, which does not reflect real-world conditions. 
For instance, creating a scenario where Karie's medicine scanner constantly fails is an unrealistic representation of the environmental scenario. 
Hence, the extent of model coverage achieved during environment simulations generated with \approach{} is deemed acceptable.

\begin{tcolorbox}[colback=teal!10!white,colframe=white!75!white, left=7pt, right=7pt, top=7pt, bottom=7pt] 
\textbf{RQ1 Finding:} Across all three devices, high and consistent model coverage was attained with the Exponential and Geometric distributions, demonstrating their applicability in simulations of uncertain environments with \approach{}. 
\end{tcolorbox}

\subsubsection{RQ2 Results}\label{sec:rq2res}

In response to RQ2 regarding the diversity of uncertain scenarios, \Cref{tab-rq2results} provides the results for the Karie, Medido, and Pilly devices for each probability distribution.

For the Karie device, a high average diversity is observed with Exponential, Geometric, and Logarithmic distributions. 
Meanwhile, the second highest average diversity is noticed in Poisson, Triangular, and Normal distributions.
Among these results, the Exponential distribution, despite its high diversity, demonstrates the lowest STD. 
This indicates that numerous diverse uncertain scenarios were generated consistently with Exponential distribution. 
On the contrary, the Uniform distribution demonstrates the lowest average diversity, implying that multiple simulations executed homogeneous types of uncertain scenarios.

In the case of the Medido device, Normal and Poisson distributions show a high average diversity, while Exponential, Logarithmic, and Triangular distributions demonstrate the second-highest average diversity. 
The STD values for Normal and Logarithmic distributions are the lowest compared to others, suggesting these distributions maintained consistency while generating many diverse uncertain scenarios across all simulations.
Conversely, the results for Uniform distribution demonstrate the lowest diversity with a slightly high STD, indicating that numerous simulations executed similar types of uncertain scenarios. 

Finally, for the Pilly device, high average diversity values were observed with Exponential, Triangular, and Logarithmic distributions. 
Additionally, Normal and Poisson distributions demonstrate the second-highest average diversity.
The STD value for Exponential distribution is the lowest among all, indicating that a significant number of diverse uncertain scenarios were consistently generated in all simulations using this distribution.
Meanwhile, Pilly's results for uniform distribution are similar to those of Karie and Medido. 
This implies that many simulations using the Uniform distribution consistently produced the same types of uncertain scenarios.

Upon examining the diversity results, it is evident that the highest diversity value approximates to 0.62. 
None of the simulations exhibited a diversity value of 1, signifying full diversity. 
The primary cause for this, besides uncertainties, is the design of state machines for various environment components, wherein multiple transitions lead to the same state. 
This results in repeated executions of identical states, thereby decreasing diversity.  
For example, in the UI failure scenario, multiple UI interactions may lead to \emph{UIFailure} state as shown in \Cref{fig:patient-ui-sm}. 
Consequently, the diversity level attained during the simulation of uncertain scenarios is reasonable.

\begin{tcolorbox}[colback=teal!10!white,colframe=white!75!white, left=7pt, right=7pt, top=7pt, bottom=7pt] 
\textbf{RQ2 Finding:} \approach{} generated environment simulations using Exponential and Logarithmic distributions consistently produced numerous diverse uncertain scenarios across all experimental runs and for most devices. 
\end{tcolorbox}

\subsection{\textcolor{black}{Discussion}}
\textcolor{black}{ 
The results for RQ1 and RQ2 demonstrated \approach{}'s applicability to three medicine dispensers by evaluating the suitability of various probability distributions in capturing the environmental uncertainties. 
In the following, we provide a detailed analysis and discuss the practical implications of these results. 
}

\textcolor{black}{
\subsubsection{Determining Probability Distributions}
Upon analyzing the results thoroughly, it was observed that determining suitable probability distribution involves additional factors, essential for applicability. 
The first factor is the feature (e.g., facial recognition) and component (e.g., camera) variation among similar types of devices, which makes a single probability distribution potentially unsuitable for all devices. 
In the case of medicine dispensers, generating an alarm for dose intake time is a common feature. 
However, the dispensing procedure differs among medicine dispensers developed by different vendors. 
For example, Karie's dispensing process involves face recognition through a camera, whereas Medido and Pilly do not. 
Consequently, Karie is equipped with a camera component, which is not present in Medido and Pilly. 
Due to the variations in the components, each device exhibits different behaviors and the associated environmental uncertainties. 
It is evident from the RQ1 and RQ2 results that Exponential distribution outperformed other distributions for all three devices. 
When considering both model coverage and scenario diversity, it was observed that Geometric distribution is well-suited for Karie but not for Medido or Pilly. 
Similarly, the Logarithmic distribution appears suitable for the Medido, while the Exponential distribution seems appropriate for the Pilly. 
}

\textcolor{black}{
The second factor is each component's unique behavior within a single device. 
During the simulation of different device components, it was noticed that each component possesses different environmental uncertainties. 
Due to this, each component within a single device may adhere to different probability distributions. 
For example, scenarios involving power failure in the battery component usually have a lower likelihood of occurrence compared to the network component. 
This is because the network component involves connectivity behavior, which is constantly active and necessary for maintaining a connection with an IoT application. 
Therefore, the probability distribution followed by the network component may differ from that of the battery component. 
Furthermore, components like the camera and scanner allow multiple attempts in case of a mismatch. 
For these components, probability distributions involving trials, such as Geometric or Poisson, may be suitable. 
Practitioners, when specifying these distributions, also need to consider the uncertain behavior of individual components. 
This consideration is important when customizing our approach for a specific device. 
Moreover, empirically analyzing probability distribution for individual components could be a potential future research direction.  
\subsubsection{Model Coverage and Diversity}
The overall results demonstrated model coverage of approximately 61\%, and uncertain scenarios diversity of value close to 0.62. 
In addition to the probabilistic simulation of environment models, we identified two aspects affecting the coverage and diversity. 
One aspect is the routine operations of devices and environmental components, which lead to repeated execution of the same states and transitions. 
For example, Karie regularly checks for a new or updated medication schedule, which leads to repeatedly executing \emph{LoadingMedsPlan} state (\Cref{fig:dev-dt-sm}). 
Similarly, in dispensing operations, which are common among medicine dispensers, Karie follows a medication plan and dispenses pills multiple times. 
This leads to multiple executions of \emph{FollowsMedsPlan} and \emph{DispensingMeds} states (\Cref{fig:dev-dt-sm}). 
Executing the same states multiple times impacts both model coverage and the diversity of uncertain scenarios. 
However, it is important to note that these are the intended operations of devices. 
}

\textcolor{black}{
Another aspect to consider is the frequency of occurrences of uncertain environmental events. 
In our experience, some uncertain events tend to occur more frequently than others in practical settings. 
For example, a medicine pack getting stuck during dispensing happens less frequently than a network connectivity issue. 
This leads to the generation of more events corresponding to network connectivity than to medicine delivery failures. 
As a result, the simulation will execute scenarios that are highly likely to occur, which ultimately reduces the overall diversity of uncertain scenarios. 
From the analysis of our experimental results, we noted similar scenarios that resemble those typically observed in real-world environments. 
}

\textcolor{black}{
For our experiments, we specified uncertainties in the environment models based on our familiarity with real devices and discussion with practitioners from Oslo City's healthcare department. 
Our experimental results demonstrated the impact of various probability distributions in terms of model coverage and diversity. 
Considering the results and implications of the evaluation, testers can customize our approach for a particular testing intent. 
For instance, if a testing intent necessitates simulating a large number of uncertain failure scenarios, a tester would need to assign high probabilities to relevant model elements, such as failure states or transitions. 
On the other hand, to balance between failure and normal scenarios, a tester can assign equal probabilities to both types of model elements. 
It is important to note that these testing-specific customizations may not accurately reflect real-world environmental scenarios. 
However, they would facilitate more rigorous testing with DTs. 
}

\textcolor{black}{
\subsubsection{Uncertainty Considerations}
Generally, uncertainty is categorized into two types: \emph{epistemic uncertainty}, which relates to the lack of knowledge about the system, and \emph{aleatory uncertainty}, which is associated with the inherent randomness of phenomena~\cite{der2009aleatory}. 
Depending on the specific system and domain, various uncertainty classifications and subcategories are also presented in the literature, e.g., for complex systems~\cite{thunnissen2003uncertainty}. 
A detailed classification by Troya et al.~\cite{troya2021uncertainty} presents six uncertainty types: \emph{measurement}, \emph{occurrence}, \emph{belief}, \emph{design}, \emph{behavior}, and \emph{spatiotemporal} uncertainty. 
Among these types, \emph{belief} uncertainty is related to a modeler/tester's \emph{degree of belief/confidence} about a particular model element, such as a class, state, or transition. 
For example, a modeler or tester may have 90\% confidence that a device will transition into a \textit{shutdown} state in the event of a power failure. 
}

\textcolor{black}{
In our experiments, we developed environment models for three different medicine dispensers. 
The uncertainties specified for these models belong to the \emph{belief} uncertainty category. 
These uncertainties were identified based on our prior experience with these devices, considering each device's functionality and scenarios necessary for testing. 
For other categories of medical devices with different environmental conditions and testing requirements, other types of uncertainties may be relevant. 
For instance, in the case of blood pressure monitoring devices, measurement uncertainty may also apply in addition to belief uncertainty. 
To illustrate, a tester may believe that a blood pressure device could be inaccurate about 5-15\% of the time, and the corresponding measurement inaccuracy could be up to 20 mm Hg (i.e., $\pm$ 20). 
In such cases, approaches for specifying measurement uncertainty in models, such as the work by Bertoia et al.~\cite{bertoa2020incorporating}, could be valuable. 
Due to the unavailability of measurement devices in our experimental setup, we could not incorporate them into our experiments. 
Nevertheless, exploring uncertainties associated with diverse types of devices could be an intriguing future direction. 
}

\subsection{Threats to Validity}
The potential threats that could impact the validity of our experiment, as well as the measures we have taken to mitigate these risks, are discussed in the following sections.

\subsubsection{Internal Validity Threat}
A potential threat to the internal validity of our experiment may arise due to the creation of environment models. 
To reduce this threat, we carefully developed these models using the documentation provided by Oslo City's healthcare department. 
Furthermore, we engaged in various discussion sessions with the technical team from Oslo City's healthcare department.

\subsubsection{External Validity Threat}
To handle external validity threats, we incorporated three medicine dispensers with diverse functionalities in our experiments. 
These dispensers span a range from basic to advanced features, and from low-end to high-end devices, thereby providing a representative sample.
Moreover, these dispensers are extensively deployed throughout Oslo City and across numerous counties in Norway, making them representative case studies drawn from real-world scenarios. 
While our experimental results may not generalize for all kinds of medical devices, this possible lack of generalizability is a common threat in most experimental research~\cite{khan2019aspectocl,sartaj2024efficient}.

\subsubsection{Construct Validity Threat}
A possible construct validity threat can occur due to an inappropriate choice of metrics and measures.
To address this threat, we carefully selected evaluation metrics and measures adhering to recommended practices. 
Specifically, we utilized model coverage and diversity measures to analyze RQ1 and RQ2 results. 
Additionally, we employed descriptive statistics, i.e., mean and standard deviation, to present a summarized overview of our experimental findings.

\subsubsection{Conclusion Validity Threat}
A potential threat to the conclusion validity in our experiment could arise from the inherent randomness in environment simulations.
To account for the randomness factor in our experiments, we repeated each experiment 30 times following established guidelines for experiments of a similar nature~\cite{arcuri2011practical}.

\section{Insights and Takeaways}\label{sec:insights}
This section presents insights derived from our modeling work and experiments, offering key takeaways for researchers and practitioners in related fields.

\subsection{Models Correctness}
Our approach requires creating models of the device's environment and DT. 
Ensuring the correctness of these models is important. 
For that, our approach depends on the tester to carefully develop and verify these models, which is a common practice~\cite{sartaj2021testing,sartaj2024automated}. 
Alternatively, automated model verification techniques (like~\cite{altoyan2023proving}) can also be applied for this purpose. In addition, model testing approaches can be applied to ensure the correctness of the models, as argued in \cite{LionelModelTesting}. However, testing models is challenging since, for instance, a test oracle that could tell what a correct model is does not exist.

\subsection{Behavior Learning}
In this work, we created models manually. Naturally, an alternative is to learn at least a partial model from data, if available. This is an interesting direction we intend to follow since we can access some operational data of devices. To this end, we aim to investigate various machine learning techniques (e.g., Hidden Markov Models, Deep Learning, and Reinforcement Learning) to learn an initial version of models later improved by domain experts manually. Moreover, automata learning techniques~\cite{aichernig2018model} are interesting to investigate to learn an initial version of behavioral models and assess their impact on the environment simulations.

\subsection{Uncertain Environment Evaluation}
We observed that assessing an uncertain simulated environment, particularly when compared to a real environment, introduces significant challenges.
First, creating uncertain scenarios for a real environment is challenging. This is mainly due to our limited knowledge of the real environment, which prevents us from creating such scenarios. Second, it is also difficult to automatically assess the realism of the created scenarios. Currently, it is a manual process, which deserves more research to find ways to automatically determine the realism of uncertain environment scenarios. Third, comparing simulated environmental scenarios with real environment scenarios is a difficult task. 
This is primarily attributed to the device's inherent operating mechanisms, which cannot be manipulated or controlled. 
For instance, making comparisons based on a scenario where medicine gets stuck during dispensing would necessitate the dispenser to enter a state of `medicine-stuck', a situation that is not controllable. 
Given the lack of control over the device's operating mechanisms, comparing the simulated and real environments is infeasible.

\subsection{Configuring Probability Distributions}
In our experiments, we used default parameter values for probability distributions. 
Nevertheless, if domain experts know specific parameter values or can extract them from a real environment, they may facilitate a more precise simulation of the environment.
For example, several probability distributions (such as Binomial and Bernoulli) require the number of trials and the probability value. 
If domain experts can determine the number of trials and the associated probabilities of success and failure based on prior knowledge or manual observation, the resulting uncertain scenarios will more accurately simulate real-world situations. 
To illustrate, if domain experts know or observe that pressing a button on a device more than 10 times results in a failure state, this number of trials can be supplied as a parameter value for the probability distributions. 
Consequently, the simulation of such an uncertain scenario would more closely resemble a real-world situation.

Moreover, to determine which probability distribution fits a particular device, historical data or domain experts' experiences regarding uncertain events are required~\cite{mun2010modeling}. 
In our case, this information was inaccessible due to the confidentiality policies of various third-party device vendors.  
If such information is available beforehand, various automated tools (like Risk Simulator~\cite{mun2015risk}) can be employed to find a probability distribution that follows the device's realistic behavior. The probability distribution determined in this way can ultimately be incorporated into our approach. 

\subsection{Industry Evaluation of Modeling Approach}
A realistic way to assess the effectiveness of modeling approaches is to perform controlled experiments with real modelers. However, this is a very costly task and requires extensive resources, which we do not have. In our context, conducting such experiments with practitioners at Oslo City is impossible. As we explained in the evaluation section, we adopted alternative ways to assess the effectiveness of our modeling approach, which indeed indicates its value.

\subsection{\textcolor{black}{\approach{} role in Testing}}
\textcolor{black}{
DTs have largely contributed to testing processes in numerous safety-critical domains, as discussed by Somer et al.~\cite{somers2023digital}.  
Our work primarily focuses on leveraging DTs and their accompanying environment simulations to support testing IoT-based healthcare applications at both system and integration levels. 
Given the safety-critical nature of such applications, an important consideration is the role of DTs and their environment simulation in testing. 
Our experiments with operating DTs of medical devices demonstrated that DTs exhibit high fidelity with their corresponding physical counterparts, and thus can substitute medical devices for testing~\cite{sartaj2023hita,sartaj2024modelbased}. 
We also observed that environmental factors of medical devices play an important role in testing. 
Without considering the environmental impact, several critical scenarios (discussed in \Cref{sec:motivation}) necessary for testing may be missed. 
Therefore, in this paper, we introduce \approach{}, an approach designed to operate DTs along with their environmental simulations. 
The ultimate aim is to facilitate more rigorous and automated test execution for testing IoT-based healthcare applications.  
}

\subsection{Approach Generalizability \textcolor{black}{and Applicability}}
Our approach primarily focused on simulating the environment for DTs of medical devices. 
While our evaluation specifically addressed three types of medicine dispensers integrated with an IoT-based healthcare application, it is worth noting that our approach applies to a broader range of medical devices for the following reasons. First, the models created for each device type can be easily adapted for later device versions with simple configurations. Second, the other existing dispensing devices, other than the ones we modeled, share the common behavior of medicine dispensing. This means that the models can be configured for other dispensing devices. Third, for future dispensing devices that do not belong to the types of devices, we could largely use the models we created since the main functionality of these devices remains the same. Nonetheless, for totally different devices, one needs to create models from scratch.     

\textcolor{black}{
Given our industrial context of a healthcare IoT application and the availability of real-world medical devices, our work mainly focused on simulating environments for DTs specific to these devices. 
However, the modeling methodology of our approach has broader applicability to other types of IoT devices. 
The environment modeling profile and events model library comprise generic concepts that potentially apply to devices in other domains.
For instance, concepts like \emph{User}, \emph{Power}, and \emph{Sensor} from our environment modeling profile might also be relevant for modeling environments for DTs of other IoT devices, such as mobile security alarm devices for emergencies. 
Moreover, the behavioral models can be customized to accommodate the specific type of device and the corresponding environmental conditions.  
Evaluating the applicability to other IoT devices requires access to them and their associated applications. 
Without access to varied types of IoT devices, our research is currently confined to medical devices. 
In addition to IoT devices, our modeling methodology also has the potential to apply in other domains, such as cyber-physical systems. 
However, we recognize the need for a dedicated future study to evaluate the applicability and generalizability of our modeling notations to other related domains. 
}

\section{Related Works}\label{sec:relatedworks}
This section contextualizes our work among the existing literature, addressing key aspects comprising digital twins in different domains, including IoT, environment modeling, and uncertainty modeling.

\subsection{Digital Twins}
In the IoT domain, Elayan et al.~\cite{elayan2021digital} utilized machine learning techniques to create DTs of Electrocardiogram (ECG) for predicting heart conditions in IoT-based healthcare systems. 
In contrast, our work contributes to modeling and simulating the environment for DTs of medical devices. 
Nguyen et al.~\cite{nguyen2022digital} presented a tool that uses DT to create a testing environment for an IoT system. 
Instead, our work focuses on creating and simulating the environment for medical devices' DTs. 
Sleuters et al.~\cite{sleuters2019digital} introduced a DT creation method for analyzing the behavior of IoT systems; however, our work targets a different objective. 
Jiang et al.~\cite{jiang2021digital} used DT to create an architecture for integrating IoT devices efficiently, whereas our work focuses on DT environment simulation. 
Kirchhof et al.~\cite{kirchhof2021understanding} employed DTs for analyzing the behavior of model-based IoT systems. 
In comparison, our work proposes a model-driven methodology for simulating the environment for DTs. 
Sciullo et al.~\cite{sciullo2024relativistic} presented a framework for generating and calibrating DTs of IoT smart devices. 
On the contrary, our work targets generating environment simulations for DTs. 
In a similar line of work, we proposed a model-based approach for creating and operating DTs of IoT medical devices~\cite{sartaj2024modelbased}. 
The research outlined in this paper builds upon our prior work, aiming to fill the gap in environment simulation for DTs. 

\textcolor{black}{
DTs are prevalent in the CPS domain, especially in smart manufacturing~\cite{somers2023digital,tao2019digital}. 
In the context of DTs for smart devices, Shoukat et al.~\cite{shoukat2024smart} presented an architecture for modeling home devices DTs to facilitate efficient healthcare for the elderly. 
Pirbhulal et al.~\cite{pirbhulal2024cognitive} proposed a framework for cognitive DTs with the ability to handle probable cyber security threats in healthcare applications. 
Zhou et al.~\cite{zhou2024toward} introduced an approach to construct DTs of motion-tracking devices used by humans in the CPS context. 
Damjanovic and Behrendt~\cite{damjanovic2019open} presented an open-source DT demonstrator for smart CPS considering DT models, microservices, and data management. 
Kirchhof et al.~\cite{kirchhof2020model} presented a domain-specific language to integrate CPS and DTs and to synchronize their development. 
Dobaj et al.~\cite{dobaj2022towards} proposed a DT-based self-adaptive model for CPS to streamline DevOps and testing processes of industrial CPS. 
For socio-technical systems, Barat et al.~\cite{barat2022digital} proposed a model-based approach that uses DTs to enable safe simulation and experimentation.   
All the aforementioned works target either DT generation or using DT for CPS and to facilitate smart manufacturing. 
In comparison, our work mainly focuses on simulating uncertain environments for DTs of IoT medical devices. 
Additionally, our work can complement existing approaches for DTs by facilitating environment simulations. 
}

Besides the IoT and CPS domains, DTs are also widely used in other fields like handling graphical user interactions with DTs~\cite{bano2022process}, linking DTs with physical twins~\cite{munoz2021using}, DTs of satellite systems~\cite{christofi2022novel}, and DTs in web applications~\cite{bonney2021digital}. 
\textcolor{black}{ 
The main differences between our work and these studies lie in the application domain and the contribution associated with DTs. 
Specifically, our work is focused on the healthcare IoT domain and the proposed approach is designed to simulate uncertain environments for DTs. 
}
In related research, Bersani et al.~\cite{bersani2022engineering} proposed an initial idea for developing a medical device (a ventilator) DT capable of dealing with security concerns and environmental uncertainties.
\textcolor{black}{
Compared to this work, first, our work mainly targets IoT medical devices, which are different from ventilator devices. 
The second main difference is that we aim to generate uncertain environmental simulations for DTs to facilitate testing, while their work focuses on augmenting DTs with reliability and trustworthiness characteristics.  
}

\subsection{Environment Modeling}
In the environment modeling aspect, Iqbal et al.~\cite{iqbal2015environment} proposed a model-based approach specifically designed for modeling and simulating the environment of real-time embedded systems. 
\textcolor{black}{
The main difference with this work is related to domain-specific behavior and uncertainties of environment components. 
For instance, components like \emph{User} and \emph{Sensor} from our environment modeling profile are common to embedded systems. 
However, a \emph{User}'s behavior and the associated uncertainties for an embedded system like a vending machine will differ from that of a \emph{User} of a medical device. 
} 
Llopis et al.~\cite{llopis2023modeling} presented a model-driven approach to design and synchronize environments, particularly for DTs of smart rooms. 
\textcolor{black}{
Similar to this, our approach focuses on environment simulation for DTs. 
However, our approach incorporates uncertainty-wise environment simulation, whereas their work does not consider uncertainties. 
Furthermore, while their modeling methodology is tailored for smart room DTs, our work specifically aims at IoT medical devices. 
} 
Haris et al.~\cite{haris2019sensyml} presented a simulation environment featuring IoT sensor simulation, data collection from real sensors, and testing at a large scale with IoT nodes. 
\textcolor{black}{
While their work targets creating a simulation environment with simulated sensors, our work focuses on simulating uncertain environments for DTs of medical devices. 
} 
Paredis et al.~\cite{paredis2021exploring} developed environment models of robot systems for creating their digital shadows. 
\textcolor{black}{
The main difference is that our work targets uncertain environment simulations for DTs, not the digital shadows. 
Furthermore, our work considers environmental uncertainties, a factor not considered in their work. 
}

\subsection{Uncertainty Modeling}
Numerous studies focus on modeling uncertainty across diverse systems~\cite{troya2021uncertainty}. 
Here, we discuss works that are closely relevant to our research. 
In research on uncertainty within complex systems, Eramo et al.~\cite{eramo2016approaching} proposed a model-based approach to manage uncertainty by configuring uncertainty models of complex systems. 
In works related to CPS, Burgue{\~n}o et al.~\cite{burgueno2018expressing} presented a proposal for specifying uncertainty in models and transformation rules of CPS. 
Similarly, Zhang et al.~\cite{zhang2019uncertainty} presented a framework that includes an uncertainty profile and modeling libraries to develop test-ready models for CPS. 
In the context of measurement uncertainty, Bertoa et al.~\cite{bertoa2018expressing,bertoa2020incorporating} introduced uncertainty data types as an extension to UML/OCL data types, specifically for modeling measurement uncertainties. 
Similarly, Burgue{\~n}o et al.~\cite{burgueno2023dealing} introduced a method featuring a UML profile for representing belief uncertainty within domain models.
In works associated with self-adaptive systems, C{\'a}mara et al.~\cite{camara2022addressing,camara2022uncertainty} outlined the challenges of modeling and analyzing uncertainty in self-adaptive systems. 
Moreover, Jezequel and Vallecillo~\cite{jezequel2023uncertainty} proposed using random variables to incorporate measurement uncertainties in models for simulating self-adaptive systems. 
Recently, Jongeling and Vallecillo~\cite{jongeling2023uncertainty} presented a method to facilitate industry practitioners in specifying uncertainties by annotating models and checking inconsistencies. 

Compared to the above-mentioned works, our work does not focus on formulating a method for modeling uncertainties. 
Instead, we aim to apply uncertainty modeling techniques to model and stimulate environmental uncertainties for DTs of IoT smart medical devices.

\section{Conclusion}\label{sec:conclusion}
Considering the importance of simulating the environment for medical devices' DTs, this paper presented \approach{}---an approach featured with environment modeling profile, environment events model library, methodology for modeling a device's environment, and uncertainty-based simulation of the environment.
We evaluated \approach{} in a real-world experimental setup supplied by Oslo City's healthcare department, comprising three medicine dispensers (Karie, Medido, and Pilly) connected to an IoT-based healthcare application. 
Our evaluation aimed to investigate the coverage of the environment models and the diversity of the uncertain scenarios generated for the medicine dispensers' DTs.
Results show that \approach{} attained approximately 61\% coverage of environment models during all simulation runs for the Karie, Medido, and Pilly devices. 
Additionally, results demonstrate that \approach{} generated diverse uncertain scenarios across multiple environmental simulations for every device. 

\backmatter

\bmhead{Acknowledgments}
This research work is a part of the WTT4Oslo project (No. 309175) funded by the Research Council of Norway. All the experiments reported in this paper are conducted in a laboratory setting of Simula Research Laboratory; therefore, they do not by any means reflect the quality of services Oslo City provides to its citizens. Moreover, these experiments do not reflect the quality of services various vendors provide to Oslo City.






\bibliography{refs}

\begin{thebibliography}{64}
\providecommand{\natexlab}[1]{#1}
\providecommand{\url}[1]{{#1}}
\providecommand{\urlprefix}{URL }
\providecommand{\doi}[1]{\url{https://doi.org/#1}}
\providecommand{\eprint}[2][]{\url{#2}}
 \bibcommenthead

\bibitem[{Aichernig et~al(2018)Aichernig, Mostowski, Mousavi, Tappler, and Taromirad}]{aichernig2018model}
Aichernig BK, Mostowski W, Mousavi MR, et~al (2018) Model learning and model-based testing. In: Machine Learning for Dynamic Software Analysis: Potentials and Limits: International Dagstuhl Seminar 16172, Dagstuhl Castle, Germany, April 24-27, 2016, Revised Papers, Springer, pp 74--100, \doi{10.1007/978-3-319-96562-8_3}

\bibitem[{Altoyan and Batory(2023)}]{altoyan2023proving}
Altoyan N, Batory D (2023) On proving the correctness of refactoring class diagrams of mde metamodels. ACM Transactions on Software Engineering and Methodology 32(2):1--42. \doi{10.1145/3549541}

\bibitem[{Arcuri and Briand(2011)}]{arcuri2011practical}
Arcuri A, Briand L (2011) A practical guide for using statistical tests to assess randomized algorithms in software engineering. In: Proceedings of the 33rd international conference on software engineering, pp 1--10, \doi{10.1145/1985793.1985795}

\bibitem[{Bano et~al(2022)Bano, Michael, Rumpe, Varga, and Weske}]{bano2022process}
Bano D, Michael J, Rumpe B, et~al (2022) Process-aware digital twin cockpit synthesis from event logs. Journal of Computer Languages 70:101121. \doi{10.1016/j.cola.2022.101121}

\bibitem[{Barat et~al(2022)Barat, Kulkarni, Clark, and Barn}]{barat2022digital}
Barat S, Kulkarni V, Clark T, et~al (2022) Digital twin as risk-free experimentation aid for techno-socio-economic systems. In: Proceedings of the 25th International Conference on Model Driven Engineering Languages and Systems, pp 66--75, \doi{10.1145/3550355.3552409}

\bibitem[{Bersani et~al(2022)Bersani, Braghin, Gargantini, Mirandola, Riccobene, and Scandurra}]{bersani2022engineering}
Bersani MM, Braghin C, Gargantini A, et~al (2022) Engineering of trust analysis-driven digital twins for a medical device. In: European Conference on Software Architecture, Springer, pp 467--482, \doi{10.1007/978-3-031-36889-9_31}

\bibitem[{Bertoa et~al(2018)Bertoa, Moreno, Barquero, Burgue{\~n}o, Troya, and Vallecillo}]{bertoa2018expressing}
Bertoa MF, Moreno N, Barquero G, et~al (2018) Expressing measurement uncertainty in {OCL/UML} datatypes. In: Modelling Foundations and Applications: 14th European Conference, ECMFA 2018, Held as Part of STAF 2018, Toulouse, France, June 26-28, 2018, Proceedings 14, Springer, pp 46--62, \doi{10.1007/978-3-319-92997-2_4}

\bibitem[{Bertoa et~al(2020)Bertoa, Burgue{\~n}o, Moreno, and Vallecillo}]{bertoa2020incorporating}
Bertoa MF, Burgue{\~n}o L, Moreno N, et~al (2020) Incorporating measurement uncertainty into {OCL/UML} primitive datatypes. Software and Systems Modeling 19(5):1163--1189. \doi{10.1007/s10270-019-00741-0}

\bibitem[{Bonney et~al(2021)Bonney, de~Angelis, Wagg, and Dal~Borgo}]{bonney2021digital}
Bonney MS, de~Angelis M, Wagg D, et~al (2021) Digital twin operational platform for connectivity and accessibility using flask python. In: 2021 ACM/IEEE International Conference on Model Driven Engineering Languages and Systems Companion (MODELS-C), IEEE, pp 237--241, \doi{10.1109/MODELS-C53483.2021.00042}

\bibitem[{Briand et~al(2016)Briand, Nejati, Sabetzadeh, and Bianculli}]{LionelModelTesting}
Briand L, Nejati S, Sabetzadeh M, et~al (2016) Testing the untestable: model testing of complex software-intensive systems. In: Proceedings of the 38th International Conference on Software Engineering Companion. Association for Computing Machinery, New York, NY, USA, ICSE '16, p 789–792, \doi{10.1145/2889160.2889212}

\bibitem[{Burgue{\~n}o et~al(2018)Burgue{\~n}o, Bertoa, Moreno, and Vallecillo}]{burgueno2018expressing}
Burgue{\~n}o L, Bertoa MF, Moreno N, et~al (2018) Expressing confidence in models and in model transformation elements. In: Proceedings of the 21th ACM/IEEE International Conference on Model Driven Engineering Languages and Systems, pp 57--66, \doi{10.1145/3239372.3239394}

\bibitem[{Burgue{\~n}o et~al(2023)Burgue{\~n}o, Munoz, Claris{\'o}, Cabot, G{\'e}rard, and Vallecillo}]{burgueno2023dealing}
Burgue{\~n}o L, Munoz P, Claris{\'o} R, et~al (2023) Dealing with belief uncertainty in domain models. ACM Transactions on Software Engineering and Methodology 32(2):1--34. \doi{10.1145/3542947}

\bibitem[{C{\'a}mara et~al(2022{\natexlab{a}})C{\'a}mara, Calinescu, Cheng, Garlan, Schmerl, Troya, and Vallecillo}]{camara2022addressing}
C{\'a}mara J, Calinescu R, Cheng BH, et~al (2022{\natexlab{a}}) Addressing the uncertainty interaction problem in software-intensive systems: Challenges and desiderata. In: Proceedings of the 25th International Conference on Model Driven Engineering Languages and Systems, pp 24--30, \doi{10.1145/3550355.3552438}

\bibitem[{C{\'a}mara et~al(2022{\natexlab{b}})C{\'a}mara, Troya, Vallecillo, Bencomo, Calinescu, Cheng, Garlan, and Schmerl}]{camara2022uncertainty}
C{\'a}mara J, Troya J, Vallecillo A, et~al (2022{\natexlab{b}}) The uncertainty interaction problem in self-adaptive systems. Software and Systems Modeling 21(4):1277--1294. \doi{10.1007/s10270-022-01037-6}

\bibitem[{Christofi and Pucel(2022)}]{christofi2022novel}
Christofi N, Pucel X (2022) A novel methodology to construct digital twin models for spacecraft operations using fault and behaviour trees. In: Proceedings of the 25th International Conference on Model Driven Engineering Languages and Systems: Companion Proceedings, pp 473--480, \doi{10.1145/3550356.3561550}

\bibitem[{Damjanovic-Behrendt and Behrendt(2019)}]{damjanovic2019open}
Damjanovic-Behrendt V, Behrendt W (2019) An open source approach to the design and implementation of digital twins for smart manufacturing. International Journal of Computer Integrated Manufacturing 32(4-5):366--384. \doi{10.1080/0951192X.2019.1599436}

\bibitem[{Der~Kiureghian and Ditlevsen(2009)}]{der2009aleatory}
Der~Kiureghian A, Ditlevsen O (2009) Aleatory or epistemic? does it matter? Structural safety 31(2):105--112. \doi{10.1016/j.strusafe.2008.06.020}

\bibitem[{Dobaj et~al(2022)Dobaj, Riel, Krug, Seidl, Macher, and Egretzberger}]{dobaj2022towards}
Dobaj J, Riel A, Krug T, et~al (2022) Towards digital twin-enabled devops for cps providing architecture-based service adaptation \& verification at runtime. In: Proceedings of the 17th Symposium on Software Engineering for Adaptive and Self-Managing Systems, pp 132--143, \doi{10.1145/3524844.3528057}

\bibitem[{Elayan et~al(2021)Elayan, Aloqaily, and Guizani}]{elayan2021digital}
Elayan H, Aloqaily M, Guizani M (2021) Digital twin for intelligent context-aware {IoT} healthcare systems. IEEE Internet of Things Journal 8(23):16749--16757. \doi{10.1109/JIOT.2021.3051158}

\bibitem[{Eramo et~al(2016)Eramo, Pierantonio, and Rosa}]{eramo2016approaching}
Eramo R, Pierantonio A, Rosa G (2016) Approaching collaborative modeling as an uncertainty reduction process. In: COMMitMDE@ MoDELS, Citeseer, pp 27--34

\bibitem[{Haris et~al(2019)Haris, Bisanovic, Wally, Rausch, Ratasich, Dustdar, Kappel, and Grosu}]{haris2019sensyml}
Haris I, Bisanovic V, Wally B, et~al (2019) Sensyml: Simulation environment for large-scale {IoT} applications. In: IECON 2019-45th Annual Conference of the IEEE Industrial Electronics Society, IEEE, pp 3024--3030, \doi{10.1109/IECON.2019.8927756}

\bibitem[{Hemmati et~al(2013)Hemmati, Arcuri, and Briand}]{hemmati2013achieving}
Hemmati H, Arcuri A, Briand L (2013) Achieving scalable model-based testing through test case diversity. ACM Transactions on Software Engineering and Methodology 22(1):1--42. \doi{10.1145/2430536.2430540}

\bibitem[{Iqbal et~al(2015)Iqbal, Arcuri, and Briand}]{iqbal2015environment}
Iqbal MZ, Arcuri A, Briand L (2015) Environment modeling and simulation for automated testing of soft real-time embedded software. Software \& Systems Modeling 14:483--524. \doi{10.1007/s10270-013-0328-6}

\bibitem[{J{\'e}z{\'e}quel and Vallecillo(2023)}]{jezequel2023uncertainty}
J{\'e}z{\'e}quel JM, Vallecillo A (2023) Uncertainty-aware simulation of adaptive systems. ACM Transactions on Modeling and Computer Simulation 33(3):1--19. \doi{10.1145/3589517}

\bibitem[{Jiang et~al(2021)Jiang, Guo, and Wang}]{jiang2021digital}
Jiang Z, Guo Y, Wang Z (2021) Digital twin to improve the virtual-real integration of industrial {IoT}. Journal of Industrial Information Integration 22:100196. \doi{10.1016/j.jii.2020.100196}

\bibitem[{Jongeling and Vallecillo(2023)}]{jongeling2023uncertainty}
Jongeling R, Vallecillo A (2023) Uncertainty-aware consistency checking in industrial settings. In: 2023 ACM/IEEE 26th International Conference on Model Driven Engineering Languages and Systems (MODELS), IEEE, pp 73--83, \doi{10.1109/MODELS58315.2023.00026}

\bibitem[{Karie(2024)}]{karie}
Karie (2024) Automatic medicine dispenser karie. \url{https://kariehealth.com/}, [Online; accessed 04-January-2024]

\bibitem[{Khan et~al(2019)Khan, Sartaj, Iqbal, Usman, and Arshad}]{khan2019aspectocl}
Khan MU, Sartaj H, Iqbal MZ, et~al (2019) {AspectOCL}: using aspects to ease maintenance of evolving constraint specification. Empirical Software Engineering 24(4):2674--2724. \doi{10.1007/s10664-019-09717-6}

\bibitem[{Kirchhof et~al(2020)Kirchhof, Michael, Rumpe, Varga, and Wortmann}]{kirchhof2020model}
Kirchhof JC, Michael J, Rumpe B, et~al (2020) Model-driven digital twin construction: synthesizing the integration of cyber-physical systems with their information systems. In: Proceedings of the 23rd ACM/IEEE International Conference on Model Driven Engineering Languages and Systems, pp 90--101, \doi{10.1145/3365438.3410941}

\bibitem[{Kirchhof et~al(2021)Kirchhof, Malcher, and Rumpe}]{kirchhof2021understanding}
Kirchhof JC, Malcher L, Rumpe B (2021) Understanding and improving model-driven {IoT} systems through accompanying digital twins. In: Proceedings of the 20th ACM SIGPLAN International Conference on Generative Programming: Concepts and Experiences, pp 197--209, \doi{10.1145/3486609.3487210}

\bibitem[{Llopis et~al(2023)Llopis, Criado, Iribarne, Mu{\~n}oz, Troya, and Vallecillo}]{llopis2023modeling}
Llopis JA, Criado J, Iribarne L, et~al (2023) Modeling and synchronizing digital twin environments. In: 2023 Annual Modeling and Simulation Conference (ANNSIM), IEEE, pp 245--257

\bibitem[{Medido(2024)}]{medido}
Medido (2024) Automatic medicine dispenser medido. \url{https://medido.com/en/}, [Online; accessed 04-January-2024]

\bibitem[{Mun(2010)}]{mun2010modeling}
Mun J (2010) Modeling risk: Applying Monte Carlo risk simulation, strategic real options, stochastic forecasting, and portfolio optimization, vol 580. John Wiley \& Sons

\bibitem[{Mun et~al(2015)Mun, CFC, and FRM}]{mun2015risk}
Mun J, CFC C, FRM M (2015) Risk simulator. Dublin, California, USA: Real Options Valuation

\bibitem[{Mu{\~n}oz et~al(2021)Mu{\~n}oz, Troya, and Vallecillo}]{munoz2021using}
Mu{\~n}oz P, Troya J, Vallecillo A (2021) Using {UML} and {OCL} models to realize high-level digital twins. In: 2021 ACM/IEEE International Conference on Model Driven Engineering Languages and Systems Companion (MODELS-C), IEEE, pp 212--220, \doi{10.1109/MODELS-C53483.2021.00037}

\bibitem[{Nguyen et~al(2022)Nguyen, Segovia, Mallouli, Oca, and Cavalli}]{nguyen2022digital}
Nguyen L, Segovia M, Mallouli W, et~al (2022) Digital twin for {IoT} environments: A testing and simulation tool. In: Quality of Information and Communications Technology: 15th International Conference, QUATIC 2022, Talavera de la Reina, Spain, September 12--14, 2022, Proceedings, Springer, pp 205--219, \doi{10.1007/978-3-031-14179-9_14}

\bibitem[{Paredis and Vangheluwe(2021)}]{paredis2021exploring}
Paredis R, Vangheluwe H (2021) Exploring a digital shadow design workflow by means of a line following robot use-case. In: 2021 Annual modeling and simulation conference (ANNSIM), IEEE, pp 1--12, \doi{10.23919/ANNSIM52504.2021.9552143}

\bibitem[{Pilly(2024)}]{pilly}
Pilly (2024) Pilly sms medicine dispenser. \url{https://responssenteret.no/responsskolen/brukere/manualer-videoer/Pilly.php}, [Online; accessed 04-January-2024]

\bibitem[{Pirbhulal et~al(2024)Pirbhulal, Chockalingam, Abie, and Lau}]{pirbhulal2024cognitive}
Pirbhulal S, Chockalingam S, Abie H, et~al (2024) Cognitive digital twins for improving security in {IT-OT} enabled healthcare applications. In: International Conference on Human-Computer Interaction, Springer, pp 153--163, \doi{10.1007/978-3-031-61382-1_10}

\bibitem[{Py4J(2009)}]{py4j}
Py4J (2009) \url{https://www.py4j.org/}, [Online; accessed 28-Februray-2024]

\bibitem[{PyEcore(2023)}]{pyecore}
PyEcore (2023) \url{https://github.com/pyecore/pyecore}, [Online; accessed 04-January-2024]

\bibitem[{PyUML2(2021)}]{pyuml}
PyUML2 (2021) \url{https://github.com/pyecore/pyuml2}, [Online; accessed 08-March-2024]

\bibitem[{Roswell et~al(2021)Roswell, Dushoff, and Winfree}]{roswell2021conceptual}
Roswell M, Dushoff J, Winfree R (2021) A conceptual guide to measuring species diversity. Oikos 130(3):321--338. \doi{10.1111/oik.07202}

\bibitem[{Sartaj(2023)}]{repo}
Sartaj H (2023) {APD-DT: A Tool to Generate and Operate Digital Twins of Medicine Dispensers}. \urlprefix\url{https://github.com/Simula-COMPLEX/WTSPublic}

\bibitem[{Sartaj et~al(2019)Sartaj, Iqbal, Jilani, and Khan}]{sartaj2019search}
Sartaj H, Iqbal MZ, Jilani AAA, et~al (2019) A search-based approach to generate {MC/DC} test data for {OCL} constraints. In: Search-Based Software Engineering: 11th International Symposium, SSBSE 2019, Tallinn, Estonia, August 31--September 1, 2019, Proceedings 11, Springer, pp 105--120, \doi{10.1007/978-3-030-27455-9_8}

\bibitem[{Sartaj et~al(2020)Sartaj, Iqbal, and Khan}]{sartaj2020cdst}
Sartaj H, Iqbal MZ, Khan MU (2020) {CDST}: A toolkit for testing cockpit display systems. In: 2020 IEEE 13th International Conference on Software Testing, Validation and Verification (ICST), IEEE, pp 436--441, \doi{10.1109/ICST46399.2020.00058}

\bibitem[{Sartaj et~al(2021)Sartaj, Iqbal, and Khan}]{sartaj2021testing}
Sartaj H, Iqbal MZ, Khan MU (2021) Testing cockpit display systems of aircraft using a model-based approach. Software and Systems Modeling 20(6):1977--2002. \doi{10.1007/s10270-020-00844-z}

\bibitem[{Sartaj et~al(2023{\natexlab{a}})Sartaj, Ali, Yue, and Gjøby}]{sartaj2023hita}
Sartaj H, Ali S, Yue T, et~al (2023{\natexlab{a}}) {HITA: An Architecture for System-level Testing of Healthcare IoT Applications}. In: European Conference on Software Architecture. Springer, Cham, pp 451--468, \doi{10.1007/978-3-031-66326-0_28}

\bibitem[{Sartaj et~al(2023{\natexlab{b}})Sartaj, Ali, Yue, and Moberg}]{sartaj2023testing}
Sartaj H, Ali S, Yue T, et~al (2023{\natexlab{b}}) {Testing Real-World Healthcare {IoT} Application: Experiences and Lessons Learned}. In: Proceedings of the 31st ACM Joint European Software Engineering Conference and Symposium on the Foundations of Software Engineering. Association for Computing Machinery, ESEC/FSE 2023, p 2044–2049, \doi{10.1145/3611643.3613888}

\bibitem[{Sartaj et~al(2024{\natexlab{a}})Sartaj, Ali, and Gjøby}]{sartaj2024digital}
Sartaj H, Ali S, Gjøby JM (2024{\natexlab{a}}) Digital twins environment simulation for testing healthcare {IoT} applications. In: Proceedings of the 48th Annual Computers, Software, and Applications Conference (COMPSAC). IEEE, COMPSAC 2024, p 900–901, \doi{10.1109/COMPSAC61105.2024.00124}

\bibitem[{Sartaj et~al(2024{\natexlab{b}})Sartaj, Ali, Yue, and Moberg}]{sartaj2024modelbased}
Sartaj H, Ali S, Yue T, et~al (2024{\natexlab{b}}) Model-based digital twins of medicine dispensers for healthcare {IoT} applications. Software: Practice and Experience 54(6):1172--1192. \doi{10.1002/spe.3311}

\bibitem[{Sartaj et~al(2024{\natexlab{c}})Sartaj, Iqbal, Jilani, and Khan}]{sartaj2024efficient}
Sartaj H, Iqbal MZ, Jilani AAA, et~al (2024{\natexlab{c}}) Efficient test data generation for {MC/DC} with {OCL} and search. arXiv preprint arXiv:240103469

\bibitem[{Sartaj et~al(2024{\natexlab{d}})Sartaj, Muqeet, Iqbal, and Khan}]{sartaj2024automated}
Sartaj H, Muqeet A, Iqbal MZ, et~al (2024{\natexlab{d}}) Automated system-level testing of unmanned aerial systems. Automated Software Engineering 31(64):1--48. \doi{10.1007/s10515-024-00462-9}

\bibitem[{Sciullo et~al(2024)Sciullo, De~Marchi, Trotta, Montori, Bononi, and Di~Felice}]{sciullo2024relativistic}
Sciullo L, De~Marchi A, Trotta A, et~al (2024) Relativistic digital twin: Bringing the {IoT} to the future. Future Generation Computer Systems 153:521--536. \doi{10.1016/j.future.2023.12.016}

\bibitem[{Shoukat et~al(2024)Shoukat, Yan, Zhang, Cheng, Raza, and Niaz}]{shoukat2024smart}
Shoukat MU, Yan L, Zhang J, et~al (2024) Smart home for enhanced healthcare: exploring human machine interface oriented digital twin model. Multimedia Tools and Applications 83(11):31297--31315. \doi{10.1007/s11042-023-16875-9}

\bibitem[{Simpson(1949)}]{simpson1949measurement}
Simpson EH (1949) Measurement of diversity. nature 163(4148):688--688. \doi{10.1038/163688a0}

\bibitem[{Sleuters et~al(2019)Sleuters, Li, Verriet, Velikova, and Doornbos}]{sleuters2019digital}
Sleuters J, Li Y, Verriet J, et~al (2019) A digital twin method for automated behavior analysis of large-scale distributed {IoT} systems. In: 2019 14th Annual Conference System of Systems Engineering (SoSE), IEEE, pp 7--12, \doi{10.1109/SYSOSE.2019.8753845}

\bibitem[{Somers et~al(2023)Somers, Douthwaite, Wagg, Walkinshaw, and Hierons}]{somers2023digital}
Somers RJ, Douthwaite JA, Wagg DJ, et~al (2023) Digital-twin-based testing for cyber--physical systems: A systematic literature review. Information and Software Technology 156:107145. \doi{10.1016/j.infsof.2022.107145}

\bibitem[{Steinberg et~al(2009)Steinberg, Budinsky, Paternostro, and Merks}]{steinberg2009emf}
Steinberg D, Budinsky F, Paternostro M, et~al (2009) EMF: Eclipse Modeling Framework 2.0, 2nd edn. Addison-Wesley Professional

\bibitem[{Tao et~al(2019)Tao, Qi, Wang, and Nee}]{tao2019digital}
Tao F, Qi Q, Wang L, et~al (2019) Digital twins and cyber--physical systems toward smart manufacturing and industry 4.0: Correlation and comparison. Engineering 5(4):653--661. \doi{10.1016/j.eng.2019.01.014}

\bibitem[{Thunnissen(2003)}]{thunnissen2003uncertainty}
Thunnissen DP (2003) Uncertainty classification for the design and development of complex systems. In: Proceedings of the 3rd Annual Predictive Methods Conference, Veros Software, Santa Ana, CA, pp 1--16

\bibitem[{Troya et~al(2021)Troya, Moreno, Bertoa, and Vallecillo}]{troya2021uncertainty}
Troya J, Moreno N, Bertoa MF, et~al (2021) Uncertainty representation in software models: a survey. Software and Systems Modeling 20(4):1183--1213. \doi{10.1007/s10270-020-00842-1}

\bibitem[{Zhang et~al(2019)Zhang, Ali, Yue, Norgren, and Okariz}]{zhang2019uncertainty}
Zhang M, Ali S, Yue T, et~al (2019) Uncertainty-wise cyber-physical system test modeling. Software \& Systems Modeling 18:1379--1418. \doi{10.1007/s10270-017-0609-6}

\bibitem[{Zhou et~al(2024)Zhou, Wang, Pang, Shen, Wang, Wu, and Yang}]{zhou2024toward}
Zhou H, Wang L, Pang G, et~al (2024) Toward human motion digital twin: A motion capture system for human-centric applications. IEEE Transactions on Automation Science and Engineering \doi{10.1109/TASE.2024.3363169}

\end{thebibliography}

\end{document}